\documentclass{emulateapj}
\shorttitle{GRB initial Lorentz factor and $\Gamma_0-E_{\gamma, {\rm iso}}$
relation} \shortauthors{Liang et al.} \slugcomment{}
\begin{document}

\title{Constraining GRB Initial Lorentz Factor with the Afterglow Onset
Feature and Discovery of a Tight $\Gamma_0-E_{\gamma, {\rm iso}}$ Correlation}
\author{En-Wei Liang\altaffilmark{1,3}, Shuang-xi Yi\altaffilmark{1}, Jin Zhang\altaffilmark{2}, Hou-Jun L\"{u} \altaffilmark{1}, Bin-Bin
Zhang\altaffilmark{3}, and Bing Zhang\altaffilmark{3}
}\altaffiltext{1}{Department of Physics, Guangxi University, Nanning 530004,
China; lew@gxu.edu.cn}\altaffiltext{2}{College of Physics and Electronic
Engineering, Guangxi Teachers Education University, Nanning, Guangxi, 530001,
China}\altaffiltext{3}{Department of Physics and Astronomy, University of
Nevada, Las Vegas, NV 89154. zhang@physics.unlv.edu}

\begin{abstract}
The onset of GRB afterglow is characterized by a smooth bump in the early
afterglow lightcurve caused by the deceleration of the gamma-ray burst (GRB)
fireball by the circumburst medium. We make an extensive search for such a
deceleration feature, either from the literature for optical lightcurves, or
from the X-ray afterglow lightcurve catalog established with the Swift/XRT.
Twenty optically selected GRBs and 12 X-ray selected GRBs are found to show the
onset signature, among which 17 optically selected GRBs and 2 X-ray-selected
GRBs have redshift measurements. We study the optical $z$-known sample by
fitting the lightcurves with a smooth broken power-law and measure the width
($w$), rising timescale ($t_r$), and decaying timescale ($t_d$) at
full-width-at-half-maximum (FWHM). Strong mutual correlations among these
timescales and with the peak time ($t_p$) are found. The optical peak
luminosity ($L_{\rm {p,O}}$) at the lightcurve bump is anti-correlated with
$t_p$ and correlated with $w$, indicating a dimmer and broader bump at a later
peak time. The ratio $t_r/t_d$ is almost universal among bursts, but the ratio
$t_r/t_p$ varies from $0.3\sim 1$. The isotropic gamma-ray energy ($E_{\gamma,
{\rm iso}}$) is tightly correlated with $L_{{\rm p,O}}$ and $t_p$ in the burst
frame. Assuming that the bumps signal the deceleration of the GRB fireballs in
a constant density medium, we calculate the initial Lorentz factor ($\Gamma_0$)
and the deceleration radius ($R_{\rm d}$) of the GRBs in the optical-selected
sample. It is found that $\Gamma_0$ are typically a few hundreds, and the
typical deceleration radius is $R_{\rm dec}\sim 10^{17}$ cm. More intriguingly,
a tight correlation between the initial Lorentz factor and the isotropic
gamma-ray energy is found, namely $\Gamma_0 \simeq 195 E_{\gamma, {\rm iso},
52}^{0.27}$ (satisfied for both the optical and X-ray $z$-known samples). This
correlation is helpful to understand GRB physics, and may serve as an indicator
of $\Gamma_0$ for other long GRBs. We find that the early bright X-rays are
usually dominated by a different component from the external shock emission,
but occasionally (for one case) an achromatic deceleration feature is observed.
Components in X-rays would contribute to the diversity of the observed X-ray
lightcurves.
\end{abstract}


\section{Introduction\label{sec:intro}}
The fireball model has been extensively employed to explain the gamma-ray burst
(GRB) phenomenon (M\'{e}sz\'{a}ros 2002; Zhang \& M\'{e}sz\'{a}ros 2004; Piran
2004). The observed prompt gamma-ray emission is explained by synchrotron (or
inverse Compton) emission from the internal shocks in an erratic, unsteady,
relativistic fireball (Rees \& M\'{e}sz\'{a}ros 1994), and broadband afterglow
emission is attributed to synchrotron emission from the external shock when the
fireball is decelerated by a circum-burst medium (M\'{e}sz\'{a}ros \& Rees
1997; Sari et al. 1998). In order to avoid the ``compactness problem" of high
energy non-thermal photons detected from GRBs, the GRB fireball is required to
move relativistically towards earth. After the initial radiation-dominated
acceleration phase, the fireball enters the matter-dominated ``coasting" phase.
The fireball keeps an approximately same Lorentz factor until it sweeps up a
considerable amount mass from the ambient medium at the so-called deceleration
radius, after which $\Gamma$ decreases with $R$ (and the observer time $t$) as
a power law. The factor Lorentz factor during the coasting phase is called
the initial Lorentz factor ($\Gamma_0$), which is a crucial parameter to
understand GRB physics, but is poorly known for most GRBs.

Three methods have been proposed to measure the initial Lorentz factor
$\Gamma_0$. The first method is to apply the ``compactness" argument to use the
high energy photon cutoff energy in the GRB prompt emission spectrum to
estimate $\Gamma_0$. Since so far no clear cutoff feature is observed, the
common practice is to use the observed maximum photon energy to set a lower
limit on $\Gamma_0$ (Fenimore et al. 1993; Woods \& Loeb 1995; Baring \&
Harding 1997; Lithwick \& Sari 2001). This method suffers several
uncertainties. First, the cutoff energy depends on both $\Gamma_0$ and the
emission radius $R_\gamma$ (Gupta \& Zhang 2008). The method therefore relies
on the assumption of $R_\gamma =\Gamma_0^2 c \delta t$, the internal shock
radius. Such an assumption is not necessarily correct. Second, the minimum
variability time scale $\delta t$ is subject to large uncertainty, since the
GRB lightcurves are chaotic and do not have a characteristic time scale.
Finally, Fermi observations (Abdo et al. 2009b,c) indicate that some GRBs have
a distinct high energy component in the GeV range, which may come from a
different emission region. A straightforward usage of the method may lead to
erroneous conclusions. The second method to measure $\Gamma_0$ is to use the
blackbody component detected in some GRB spectra (Pe'er et al. 2007). The
limitation of the method is the difficulty of identifying blackbody components
from the GRB spectra. So far the only case that this method is securely applied
is GRB 090902B (Ryde et al. 2009). The third, but most commonly adopted, method
is to use the early afterglow lightcurve. The fireball model predicts that at
the onset of deceleration, the emission from the forward shock would display a
smooth bump in the lightcurve, the peak of which corresponds to the epoch when
essentially half of the fireball energy is transferred to the medium (Sari \&
Piran 1999; Kobayashi \& Zhang 2007). The most relevant case is the thin shell
regime, defined by that the thickness of the fireball shell satisfies
$\Delta<(E/nm_pc^2)^{1/3}\Gamma_0^{-8/3}$, where $E$ is the kinetic energy of
the fireball, $n$ is the medium density, $m_p$ the mass of proton, $c$ speed of
light (Kobayashi et al. 1999; Sari \& Piran 1999). Within this regime, the
deceleration time (the peak time at the lightcurve bump, $t_p \propto
\Gamma_0^{-8/3} (E/n)^{1/3}$, M\'esz\'aros \& Rees 1993) is sensitive depends
on the initial Lorentz factor but insensitive to other parameters. The
detection of this time can be then used to infer $\Gamma_0$. In the optical
band, early emission may be contaminated by the emission from the reverse shock
(M\'{e}sz\'{a}ros \& Rees 1997; Sari \& Piran 1999; Kobayashi 2000; Zhang et
al. 2003). However, under certain conditions (either a Poynting flux dominated
flow, Zhang \& Kobayashi 2005; or a relatively low typical synchrotron
frequency in the reverse shock, Jin \& Fan 2007), the reverse shock component
would not show up in the optical band. In these bursts, a smooth onset bump can
be detected, which signals the deceleration feature of the fireball, and hence,
can be used to constrain the initial Lorentz factor and the deceleration radius
(Sari \ Piran 1999, Zhang et al. 2003; Molinari et al. 2007;  Xue et al. 2009;
Zou \& Piran 2009).

In this paper, we constrain $\Gamma_0$ with the early GRB afterglows that show
the deceleration signature, and investigate the possible correlations among
deceleration parameters (including $\Gamma_0$) as well as the prompt gamma-ray
emission properties. We make an extensive search for the onset of afterglow
signature in the optical and X-ray lightcurves. Our sample selection is
presented in Section 2. The temporal characteristics and their correlations are
presented in Section 3, and the relation between the prompt gamma-ray
properties and the deceleration properties are investigated in Section 4. In
particular, we constrain $\Gamma_0$ and the deceleration radius of the fireball
for the $z$-known sample, and discover a tight correlation between $\Gamma_0$
and $E_{\gamma, {\rm iso}}$. Discussion and conclusions are presented in
Sections 5 and 6, respectively. A concordance cosmology with parameters $H_0 =
71$ km s$^{-1}$ Mpc$^{-1}$, $\Omega_M=0.30$, and $\Omega_{\Lambda}=0.70$ are
adopted. Notation $Q_n$ denotes $Q/10^n$ in the cgs units throughout the paper.

\section{Sample Selection and Lightcurve Fitting \label{sec:data}}
We make an extensive search for the smooth ``bump" feature as the onset of GRB
afterglow. The flares or the reversed shock emission component introduce
confusion to identify the onset feature, especially in the X-ray band. We
therefore employ the following two criteria to selection the samples. First,
the bump is smooth without superimposing significant flares around the bump.
Second, the decay slopes post the peak time is shallower than $-2$. Through
literature search, we obtain 20 optical-selected GRBs that show the afterglow
onset feature. We also go through the Swift XRT lightcurve archive that has
been processed by our group in the past (from January 2005 to September 2009),
and identify 12 X-ray selected sample with afterglow onset signature. The
observational results of these bursts are summarized in Tables 1 and 2. The
details of XRT data reduction have been presented in a series of papers
published by our group (Zhang et al. 2007a, Paper I; Liang et al. 2007, 2008,
2009; Papers II, III, IV). The prompt gamma-ray properties of the bursts are
taken from the published papers or GCN reports.

The afterglow lightcurves of the optical-selected and the X-ray-selected
samples are presented in Figures 1 and 2, respectively. The early X-ray
afterglow lightcurves for the optical-selected sample, and the optical
afterglow lightcurves for the X-ray selected sample are also present, if they
are available. In the X-ray selected sample, only few cases have simultaneous
optical observation. Nineteen out of 20 GRBs in the optical-selected sample
have simultaneously X-ray observations. We find only one case, GRB 080319C,
that shows a clear achromatic onset bump in both the optical and X-ray bands.
This suggests that the external shock emission indeed contributes to both the
optical and the X-ray band. For most optical-selected sample GRBs, on the other
hand, the early X-ray lightcurves either show erratic X-ray flares or internal
plateaus that are believed to be powered by the GRB central engine, or the
X-ray observations started only after the optical bump peak. Inspecting the two
samples shown in Figures 1 and 2, we find that the onset bumps in the
optical-selected sample are usually smoother than those observed in the
X-ray-selected sample. Seventeen out of the 20 GRBs in the optical-selected
sample and two out of 12 GRBs in the X-ray-selected sample have redshift
measurements. In the following, we mostly use only the $z$-known
optical-selected sample in our analysis but will use the $z$-known
X-ray-selected sample to confirm the findings.

We fit the lightcurves with an empirical model proposed by Kocevski \& Liang
(2001),
\begin{equation}
F(t)=F_{p}(\frac{t+t_{0}}{t_{p}+t_{0}})^{r}[\frac{d}{r+d}+\frac{r}{r+d}(\frac{t+t_{0}}{t_{p}+t_{0}})^{r+1}]^{-\frac{r+d}{r+1}},
\end{equation}
where $F_p$ is the maximum flux at \textit{$t_p$}, \textit{$t_0$} is a reference
time, and \textit{r} and {\em  d} are the rising and decaying power-law indices,
respectively. An IDL routine called {\em mpfitfun.pro} is employed for our
fitting. This routine performs Levenberg-Marquardt least-square fit to the
data for a given model. It optimizes the model parameters so that the sum of
the squares of the deviations between the data and the model becomes minimal.
The time interval and the fitting curve for each GRB are shown in Figures 1 and
2, and the fitting parameters are summarized in Tables 1 and 2 \footnote{The
reduced $\chi^2$ of our fits for some optical lightcurves are large. This is due
to the fluctuations in the lightcurves and small observational errors in the
optical data.}. Notice that the lightcurves of some GRBs, such as 050820A,
060607A, 070411, 071031, 080330, show significant energy injection or
re-brightening features after the deceleration bump. We make our fits only
around the bump. Our fits are also shown in Figures 1 and 2.

\section{Characteristics of the Onset Bump and Their Correlations}
For each fitting lightcurve, we take the full-width-at-half-maximum (FWHM) as
a characteristic width ($w$) of the bump, and measure the rising and
decaying timescales ($t_r$ and $t_d$) at FWHM. We also derive the ratios of
$t_r/t_p$ and $t_r/t_d$. The results are reported in Tables 1 and 2. The
distributions of $r$, $d$, $t_p$,$t_r$, $t_d$, $F_p$, $w$, $t_r/t_p$, and
$t_r/t_d$ are shown in Figure \ref{Dis_time}.

The following statistics applies to the optical-selected sample.
As shown in Figure \ref{Dis_time}, the rising index $r$ of most bursts are
in the range of $1-2$, with three exceptional cases, i.e., GRBs 080330
($r=0.34\pm 0.03$), 060607A ($r=4.15\pm 0.22$),
and 050820A ($4.45\pm 0.76$). Inspecting the optical lightcurve of GRB 080330,
the optical flux increases slowly, keeping almost a constant in $300-1000$
seconds post the GRB trigger. This feature is similar to that observed in GRB
060614. For GRBs 060607A, and 050820A, their optical lightcurves show a similar
behavior, with rapid increase prior to the peak and a normal decay post the
peak. The decaying index $d$ is distributed in the range of $0.44- 1.77$, with
a mean value $1.16\pm 0.34$. Except for GRBs 080330 and 061007, the decay
indices are well consistent with the isotropic forward shock models in a
constant density medium. The decay indices of the two exceptions are $\sim
1.7$, slightly steeper than the normal decay slope predicted by the isotropic
forward shock models. The $t_p$ is in the range of $10^2 - 10^3$ seconds with a
median value $\sim 380$ seconds. The distribution of $F_p$ ranges in $10^{-13}
- 10^{-8}$ ergs cm$^{-2}$ s$^{-1}$, with a mean $7.25\times 10^{-12}$ erg
cm$^{-2}$ s$^{-1}$. The width $w$ is distributed in $10^2-10^3$ seconds. The
$t_r$ and $t_d$ peak around $10^2$ seconds and $10^3$ seconds, respectively.
The ratio $t_r/t_d$ is narrowly distributed in $0.1 - 0.3$. However, the
distribution of the ratio $t_r/t_p$ is much wider, ranging in $0.3 - 1$.

We show various correlations among the deceleration parameters of the optical
selected sample in Fig. \ref{Corr_Time}, and summarize the linear correlation
coefficients from pair Spearman correlation analysis in Table 3. Tight
correlations are found among $t_r$, $t_d$, $t_p$ and $w$. The coefficients of
these correlations are larger than $0.93$. These correlations are
\begin{eqnarray}
\log t_d=(0.48\pm 0.13)+(1.06\pm 0.06) \log t_r\\
\log t_d=(-0.09\pm 0.29)+(1.17\pm 0.11) \log t_{p}\\
\log t_r=(-0.54\pm 0.22)+(1.11\pm 0.08) \log t_{p}\\
\log w=(0.05\pm 0.27)+(1.16\pm 0.10) \log t_{p}\\
\log w=(0.61\pm 0.11)+(1.05\pm 0.05) \log t_r\\
\log w=(0.15\pm 0.02)+(0.98\pm 0.01) \log t_d.
\end{eqnarray}

The tight $t_r-t_d$ correlations suggest that the structure of the bumps among
bursts are similar, indicating a universal physical origin. No correlation of
the decay index $d$ with the other parameters is found. This is consistent with
the expectation of blastwave model, in which the decay slope is dictated by the
density profile and the electron spectral index $p$ but has nothing to do with
the afterglow onset details. On the other hand, the rising index $r$ is tightly
anti-correlated with both the ratio $t_r/t_d$ and $t_r/t_p$, although it is not
correlated with $t_r$ and $t_d$. These correlations read
\begin{eqnarray}
\log r=(-0.21\pm 0.06)-(1.68\pm 0.19) \log t_r/t_p\\
\log r=(-0.15\pm 0.02)-(0.48\pm 0.05) \log t_r/t_d.
\end{eqnarray}
The $w-t_p$ correlation indicates that the wider bumps tend to peak at a later
time. In addition, both $w$ and $t_p$ in the burst frame are anti-correlated
with the peak luminosity at the bump peak $L_{\rm p,O}$,
\begin{eqnarray}
\log L_{\rm p,O}=(54.6\pm 0.8)-(2.48\pm 0.38) \log t^{'}_{p}\\
\log L_{\rm p,O}=(0.82\pm 0.79)-(2.16\pm 0.31) \log w^{'}.
\end{eqnarray}
These results suggest that a dimmer bump tends to peak at a later time with
a longer duration.

\section{Initial Lorentz factor constraints and the $\Gamma_0-E_{\gamma, {\rm iso}}$ correction}
The observation properties of the early optical bumps seem to be consistent
with being due to the onset of the external shock afterglow in the thin shell
regime. To test this hypothesis, we check the correlations of $L_{p,O}$ and the
cosmic-frame peak time $t_{p,z}=t_p/(1+z)$ with the isotropic gamma-ray energy
$E_{\gamma, {\rm iso}}$ in Figure \ref{Corr_Eiso_Lp}. We find that they are
strongly correlated, i.e.
\begin{eqnarray}
\log L_{\rm p,O}=(47.10\pm 0.12)+(1.17\pm 0.13) \log E_{\gamma, {\rm iso}, 52}\\
\log t_{\rm p,z}=(2.25\pm 0.07)-(0.38\pm 0.07) \log E_{\gamma, {\rm iso}, 52}
\end{eqnarray}
with a correlation coefficient $\kappa=0.88$ and chance probability $<10^{-4}$
for both correlations. We also derive the isotropic optical energy release from
$10$ seconds to $10^5$ seconds post the GRB trigger ($E_{\rm O,iso}$) using the
fitting curves, and show the correlation between $E_{\rm O,iso}$ and $E_{\rm
\gamma,iso}$ in Figure \ref{Corr_Eiso_Lp}. This correlation, i.e.
\begin{equation}
\log E_{\rm p,O}=(49.63\pm 0.09)+(0.74\pm 0.10) \log E_{\rm iso,52}
\end{equation}
has a correlation coefficient $\kappa=0.89$ and chance probability $<10^{-4}$.

These tight correlations indicate that a GRB with a larger $E_{\gamma, {\rm
iso}}$ tends to have a brighter optical afterglow peaking at an earlier time.
This is well consistent with the afterglow onset theory. The shape of the
lightcurve is consistent with the thin shell case (cf. the thick shell case,
see Kobayashi et al. 1999; Kobayashi \& Zhang 2007). We therefore apply the
standard afterglow model in a constant density medium (Sari \& Piran 1999) to
derive the initial Lorentz factor
\begin{equation}\label{Gamma_0}
\Gamma_0=2\left[\frac{3E_{\rm \gamma,\ iso}}{32\pi nm_pc^5 \eta t_{\rm p,z}^3}
\right]^{1/8}\simeq 193(n\eta)^{-1/8}\times \left(\frac{E_{\rm
\gamma,iso,52}}{t_{\rm p,z,2}^3} \right)^{1/8},
\end{equation}
and the deceleration radius
\begin{equation}
R_{\rm dec}=2c t_p \Gamma_p^2/(1+z)=2.25\times 10^{16}{\rm cm}\ \Gamma_{0,2}^2
t_{p,z,2},
\end{equation}
where $\eta = E_{\gamma, {\rm iso}} / E_{\rm K,iso}$ is the ratio between the
isotropic gamma-ray energy and the isotropic blastwave kinetic energy. The
results are rather insensitive to $n$ and $\eta$. In the following analysis, we
take $n=1$ cm$^{-3}$ and $\eta=0.2$ throughout.

With the data reported in Table 1, we calculate $\Gamma_0$ and $R_{\rm dec}$
for the GRBs in the optical-selected sample. The results also reported in Table
4. The distributions of the derived $\Gamma_0$ and $R_{\rm dec}$ are displayed
in Fig. \ref{Dis_Gamma}. We can see that $\Gamma_0$ is typically a few
hundreds, and the typical deceleration radius is $R_{\rm dec}=1 \times
10^{17}$. From Eq. \ref{Gamma_0}, we find $\Gamma_0$ depends on both $t_{p,z}$
and $E_{\rm iso}$. As $t_{p,z}$ is tightly correlated with $E_{\rm iso}$, one
expects tight relations between $\Gamma_0$ and $E_{\rm \gamma,iso}$ and between
$\Gamma_0$ and $t_{p,z}$. We show the two relations in Figure \ref{Corr_Gamma}.
The best fits give
\begin{equation}\label{Gamma_0-tp}
\log \Gamma_{0}=(3.69\pm 0.09)-(0.63\pm 0.04) \log t_{\rm p,z}
\end{equation}
with the correlation coefficient $\kappa=-0.97$ (chance probability $p<10^{-4}$) and
\begin{equation}\label{Gamma_0-Eiso}
\log \Gamma_0=(2.291\pm 0.002)+(0.269\pm 0.002) \log E_{\rm \gamma,iso,52}
\end{equation}
with correlation coefficient $\kappa=0.91$ (chance probability $<10^{-4}$).
These tight correlations suggest that $t_{p,z}$ and $E_{\gamma, {\rm iso}}$ are
good indicators of $\Gamma_0$ (hence $R_{\rm dec}$). In particular, the latter
correlation can be translated to
\begin{equation}
\Gamma_0 \simeq 195 E_{\gamma, {\rm iso}, 52}^{0.27}, \label{Gam-Eiso}
\end{equation}
which can be very useful to understand GRB physics (see below).

\section{Discussion}
\subsection{Implications for $\Gamma_0$ measurements and the $\Gamma_0-E_{\gamma, {\rm iso}}$ correlation}
Using a sample of GRBs that show the afterglow onset feature in the early
optical/X-ray afterglow lightcurves, we manage to constrain $\Gamma_0$ for a
good sample of GRBs which can be used to perform a statistical study of
$\Gamma_0$ for the first time. Using a different method (the opacity
constraint), the Fermi team recently sets the lower limits of $\Gamma_0$ for a
number of bright LAT GRBs, e.g. $\Gamma_{0,\min} \sim 800$ for GRB 080916C
(Abdo et al. 2009a), $\Gamma_{0,\min} \sim 1200$ for GRB 090510 (Abdo et al.
2009b), and $\Gamma_{0,\min} \sim 1000$ for GRB 090912B (Abdo et al. 2009c).
For GRB 080916C, one has $E_{\gamma, {\rm iso}} \simeq 8.8\times 10^{54}$ erg
(Abdo et al. 2009a), which corresponds to $\Gamma_0\sim 1210$ according to the
$\Gamma_0-E_{\gamma, {\rm iso}}$ relation (Eq.[\ref{Gam-Eiso}]). This is
consistent with the lower limit derived by Abdo et al. (2009a). For GRB
090902B, one has $E_{\gamma, {\rm iso}} \simeq 3.63\times 10^{54}$ erg (Abdo et
al. 2009c), which corresponds to $\Gamma_0 \sim 960$, which is smaller than
$\Gamma_{0,\min}$ derived by the Fermi team. However, as discussed in the
introduction, we believe that this is due to that the Fermi team did not use
the most conservative method to derive $\Gamma_{0,\min}$. Since GRB 090902B
clearly shows a distinct non-thermal component extending to high energy which
clearly has a different origin from the MeV component (and very likely from
different emission regions, A. Pe'er et al. 2010, in preparation). One should
have used the maximum photon energy of the MeV component to estimate
$\Gamma_{0,\min}$, which would lead to a consistent result with the prediction
of the $\Gamma_0 - E_{\gamma, {\rm iso}}$ relation. For the LAT short GRB
090510, one has $E_{\gamma, {\rm iso}}=3.5 \times 10^{52}$, which corresponds
to $\Gamma_0 \sim 270$ from the $\Gamma_0-E_{\gamma, {\rm iso}}$ relation. This
is significantly smaller than $\Gamma_{0,\min}=1200$ set by the 31 GeV photon
detected during the first second. Assuming an external shock origin of the GeV
emission, Ghirlanda et al. (2009) even derived $\Gamma_0\sim 2000$ for this
burst. The inconsistency is significant. This can be again due to the
non-conservative approach of the Fermi team (since the 31 GeV photon is from a
different component), but more probably it could be that short GRBs do not
satisfy the $\Gamma_0-E_{\gamma, {\rm iso}}$ relation derived for long GRBs.

With $\Gamma_0=100 - 1200$ derived from the $\Gamma_0-E_{iso,\gamma}$
relation, we expect that the corresponding observed $t_p$ is in the range
of $30\sim 1400$ seconds for a typical redshift $z=2$ according to the
$\Gamma_0-t_{p,z}$ correlation (Eq.[\ref{Gamma_0-tp}]).
In this time period, the observed X-rays are generally dominated by the GRB tail
emission or flares. This may be, at least partially,
the reason why not many early X-ray lightcurves show the clear afterglow
onset bump signature.

The $\Gamma_0-E_{\gamma, {\rm iso}}$ relation is very useful to pin down the
prompt emission physics of GRBs. One interesting observational correlation is
the Amati relation, i.e. $E_p\propto E_{\gamma, {\rm iso}}^{\kappa}$ (or $E_p
\propto L_{\gamma,iso}^{\kappa}$, with $\kappa \sim (0.4-0.5)$, both as a bulk
correlation among bursts and an internal correlation within a burst (Amati et
al. 2002; Wei et al. 2003; Liang et al. 2004; Yonetoku et al. 2004; Lu \& Liang
2009). However, all the prompt GRB emission models predict $E_p$ as a function
of both $E_{\gamma, {\rm iso}}$ (or $L_{\gamma,iso}$) and $\Gamma_0$ (e.g.
Table 1 of Zhang \& M\'{e}sz\'{a}ros 2002). There is no straightforward theory
that predicts the relationship between $\Gamma_0$ and $E_{\gamma, {\rm iso}}$
(or $L_{\gamma,iso}$). As a result, any theoretical model can be argued to
interpret the Amati relation, given a designed $\Gamma_0 - E_{\gamma, {\rm
iso}}$ correlation. The $\Gamma_0 - E_{\gamma, {\rm iso}}$ correlation
discovered here therefore poses great constraints on many prompt emission
models. For example, the internal shock synchrotron model predicts $E_p \propto
L^{1/2} \Gamma_0^{-2}$ (Zhang \& M\'{e}sz\'{a}ros 2002). The Amati relation
essentially requires that $\Gamma_0 \propto L^0$. The $E_p \propto E_{\gamma,
{\rm iso}}^{0.27}$ relation, combined with the trivial proportionality
$L\propto E_{\gamma, {\rm iso}}$ (a non-correlation between GRB duration and
luminosity), would lead to $E_p \propto E_{\gamma, {\rm iso}}^{1/2} E_{\gamma,
{\rm iso}}^{-0.54} \propto E_{\gamma, {\rm iso}}^{-0.04}$, which means that
$E_p$ is essentially constant for different $E_{\gamma, {\rm iso}}$ values.
This is in contradiction with the Amati relation, which can be regarded as
another argument against the internal shock synchrotron emission model of GRB
prompt emission (see also Kumar \& McMahon 2008; Zhang \& Pe'er 2009).
\subsection{Early Optical vs. X-ray emission: different physical origins?}
The simultaneous observations in the optical and X-ray bands during the early
afterglow phase also hold the key to address whether the broad band emission is
from the same emission component. With the prompt slewing capability, the X-ray
Telescope (XRT, Burrows et al. 2004) on board Swift has established a large
sample of X-ray afterglow of GRBs. Generally, the XRT lightcurves are composed
of a few power-law decaying segments and some erratic flares. Although the
X-ray lightcurves are diverse among bursts, they can be roughly classified into
three groups with a large, uniform sample established by XRT. The majority is
the so-called canonical XRT lightcurves characterized by a
steep-shallow-normal-steep decay pattern (Zhang et al. 2006; Nousek et al.
2006; O'Brien et al. 2006), although not all the segments show up in every
burst (Evans et al. 2009). The second group is composed of those lightcurves
that show a single power-law decay from early to late epochs (Liang et al.
2009; Evans et al. 2009). The third group includes some GRBs that show an
"internal plateau" that is followed by a rapid drop with a decay slope steeped
than -3 (Liang et al. 2007; Troja et al. 2007; Lyons et al. 2009). The physics
origin of the X-ray emission is still a mystery (e.g. Zhang 2007 for a review),
although the consensus is that there might be diverse origins (Liang et al.
2007). It is clear that the X-ray flares and internal plateaus are of an
internal origin. However the origin of the canonical lightcurve is still
subject to debate. ``Closure"-relation analyses suggest that the normal decay
segment following the shallow decay one in the canonical XRT lightcurves are
roughly consistent with the forward shock models (Willingale et al. 2007; Liang
et al. 2007), favoring the long lasting energy injection models for the
shallow-decay segment. However, the optical/X-ray chromatic behavior around the
shallow and normal decay transition time (Panaitescu et al. 2006; Fan \& Piran
2006; Liang et al. 2007) suggests that the X-ray and optical emissions are two
independent components. Some models attribute the entire X-ray emission to the
late emission from the central engine, probably related to the long-term
accretion history of the central engine (Ghisellini et al. 2007; Kumar et al.
2008; Cui et al. 2009; Cannizzo \& Gehrels 2009), but the consistency with the
"closure-relation" of the external shock model predictions is not naturally
explained in these models. Interestingly, Yamazaki (2009) recently suggested
that the X-ray emission might be an independent component prior to the GRB
trigger, and that the apparent shallow-to-normal transition is merely reference
time effect. Liang et al. (2009) systematically studied the Swift canonical GRB
lightcurves and confirm that shifting the reference time can indeed stretch the
canonical lightcurves to single power law lightcurves, and proposed a unified
picture for the physical origin of both the canonical and the single power-law
decaying XRT lightcurves.

As shown in Figure 1, the optical and X-ray lightcurve behaviors are
dramatically different at the early epoch, i.e., $t<1000$ seconds post the GRB
trigger. This strongly suggests that the radiations from the two energy bands
are not from the same component. The tight correlations between the observables
of the prompt gamma-rays and early optical afterglows shown in Fig.
\ref{Corr_Eiso_Lp} strongly suggest that the optical emission is likely the
``afterglow" of the GRB fireball (external shock component). This is consistent
with the smooth afterglow onset feature observed in the optical band for these
GRBs. On the other hand, it also suggests that the early bright X-ray emission
is not from the external shock. One natural question would be: where is the
external shock X-ray component? Inspecting the details of the X-ray and optical
afterglow lightcurves in Figure 1, we can see that this component is very
likely hidden underneath some brighter X-ray emission components in the early
epochs (e.g. flares, internal plateaus, or even normal plateaus).
Interestingly, one can find an X-ray decay slope similar to that of optical in
the late epochs in half of GRBs in our sample, including GRBs 050820A, 060418,
060605, 061007, 070318, 070411, 071031, 080319C, 080810, 081203, and 090102. We
therefore cannot exclude the possibility that the late X-ray and optical
emissions share the same external shock origin.

The mixture of different emission components in the X-ray observations
(Willingale et al. 2007; Liang et al. 2007; Liang et al. 2009; Nardini et al.
2009) also naturally interprets the fact that the rarity of the onset afterglow
feature observed in the X-ray band. This requires that other early X-ray
components are not bright enough to outshine the external shock component. In
fact, we find only 13 out of $\sim 400$ cases in the current XRT lightcurve
sample. They are shown in Figure 2. Since the optical band is less affected by
the other emission components related to the central engine, one naively
expects that the X-ray onset cases should have achromatic optical onset feature
as well. Unfortunately, the optical/X-ray joint observations in this sample are
rare: only GRB 080507 and GRB 080319C have early optical observations. The
optical lightcurve of GRB 080507 is sparse. For GRB 080319C one indeed observes
an achromatic onset feature. There is an earlier decay feature in the optical
lightcurve of GRB 080319C. It may be associated with an internal emission
component within such an interpretation.

\section{Conclusions}
We have made an extensively search for the afterglow onset ``bump" feature from
early afterglow lightcurves, both in the optical band (through literature
survey) and in the X-ray band (through systematically analyzing the Swift XRT
data). Twenty GRBs are identified in the optical-selected sample and 12 GRBs
are found in the X-ray-selected sample. We fit the onset bumps with a smooth
broken power-law and measure their characteristics. The rising index $r$ for
most bursts is $1-2$, and the decay index $d$ is $0.44- 1.77$. These are well
consistent with the forward shock models. The peak time $t_p$ is in $10^2-
10^3$ seconds with a median value of $\sim 380$ seconds. The width of the bumps
measured at FWHM is $10^2 - 10^3$ seconds, and the typical rising time $t_r$
and decaying time $t_d$ are $10^2$ seconds and $10^3$ seconds, respectively.
The ratio of $t_r/t_d$ is narrowly distribution around $0.1- 0.3$, and the
ratio $t_r/t_p$ has a distribution in the range of $0.3-1$.

Most GRBs in our optical-selected sample have redshift measurements. We analyze
pair correlations among the bump characteristics. We find that the
pulse width, rising time, decaying time, and the peak time are strongly
correlated. Bumps that peak later are dimmer and wider. No correlation between
the decay index $d$ with other parameters is found, but the rising index $r$
is tightly anti-correlated with both the ratio $t_r/t_d$ and $t_r/t_p$, although
it is not correlated with $t_r$ and $t_d$.

We analyze the relation of the optical afterglow bumps with prompt gamma-ray
properties. We find that a GRB with larger $E_{\gamma, {\rm iso}}$ tends to
have a brighter optical afterglow, and tends to be decelerated by the
surrounding medium earlier.  These tight correlations strongly suggests an
external shock afterglow origin of the early optical emission.

Within the framework of the standard forward model in a constant density
circumburst medium, we calculate the initial Lorentz factor $\Gamma_0$ and
the deceleration radius $R_{\rm dec}$ for the GRBs in the optical-selected
sample. The derived $\Gamma_0$ ranges from 100 to about 600, while $R_{\rm dec}$
narrowly distributed around $10^{17}$ cm.

Intriguingly, we discover a tight correlation between $\Gamma_0$ and
$E_{\gamma, {\rm iso}}$. For typical values $n=1$ cm$^{-3}$ and $\eta=0.2$
($\Gamma_0$ is insensitive to the values of $n$ and $\eta$, we obtain the
correlation Eq.(\ref{Gam-Eiso}). This correlation is very important to
understand GRB prompt emission physics, and may serve as an indicator of
$\Gamma_0$ for other long GRBs. In particular, the correlation disfavors the
internal shock synchrotron emission model for the GRB prompt emission. The
extrapolation of the correlation is consistent with the $\Gamma_{0,\min}$ of
GRB 080916C derived by the Fermi team (Abdo et al. 2009a), but are inconsistent
with the $\Gamma_{0,\min}$ of GRB 090510 and GRB 090902B derived by the Fermi
team (Abdo et al. 2009b,c). We point out that this is because the Fermi team
did not use a more conservative approach to set $\Gamma_{0,\min}$.

The X-ray-selected sample only has two cases with redshift measurements. The
derived $\Gamma_0$'s from these two cases are also consistent with the
$\Gamma_0-E_{\gamma, {\rm iso}}$ correlation derived from the optical-selected
sample. There is one case (GRB 080319C) that shows an achromatic afterglow
onset feature.

Most optical-selected sample has early X-ray emission components not from the
external shock. This reinforces the diverse origin of early X-ray afterglow in
most GRBs.

\acknowledgments We acknowledge the use of the public data from the Swift data
archive. This work is supported by the National Natural Science Foundation of
China under grants No. 10873002, the National Basic Research Program (''973"
Program) of China (Grant 2009CB824800), the research foundation of Guangxi
University (M30520). It is also partially supported by NASA NNX09AT66G and
NNX10AD48G, as well as NSF AST-0908362. BBZ acknowledges the President's Fellow
Ship and GPSA awards from UNLV.

\clearpage
\clearpage
\begin{deluxetable}{lccccccccccccc}

\tablewidth{500pt} \tabletypesize{\tiny}
\tablecaption{Optical observations and fitting result for our sample}
\tablenum{1}

\tablehead{ \colhead{GRB(Band)}& \colhead{$z^{Ref}$}&
\colhead{$F_{m}$\tablenotemark{a}}
&\colhead{$t_{p}$\tablenotemark{a}}&\colhead{$\alpha_1$}&
\colhead{$\alpha_2$}&\colhead{$\chi^2$/dof}
&\colhead{$w$\tablenotemark{a}}&\colhead{$t_r$\tablenotemark{a}}&\colhead{$t_d$\tablenotemark{a}}&\colhead{$t_r/t_d$}&\colhead{$t_r/t_p$}&\colhead{Refs.}}

\startdata
030418(V) &...           &2.51$\pm$0.06 &1344.5$\pm$78.6 &1.17$\pm$0.20 &0.74$\pm$0.08 &26/9  &5816&953&4863&0.20&0.71&(18)&\\
050730(V) &$3.97^{(1)}$  &4.00$\pm$0.45  &590.7$\pm$131.5 &1.36$\pm$0.43 &1.02$\pm$0.15 &25/7  &1836&339&1497&0.23&0.57&(19)&\\
050820A(R)&$2.615^{(2)}$ &21.10$\pm$1.25 &391.0$\pm$16.7  &4.45$\pm$0.76 &1.04$\pm$0.01 &47/7  &711&120&592&0.20&0.31&(20)&\\
060418(H) &$1.49^{(3)}$  &88.70$\pm$1.71 &153.3$\pm$3.3   &2.70$\pm$0.22 &1.27$\pm$0.02 &16/8  &298&65&233&0.28&0.42&(21)&\\
060605(R) &$3.8^{(4)}$   &9.59$\pm$0.20 &399.1$\pm$13.0  &0.90$\pm$0.09 &1.17$\pm$0.05 &74/50 &1313&281&1032&0.27&0.70&(22)&\\
060607A(H)&$3.082^{(5)}$ &28.60$\pm$0.49 &180.9$\pm$2.4   &4.15$\pm$0.22 &1.32$\pm$0.04 &46/23 &259&57&202&0.28&0.31&(21)&\\
060904B(V)&$0.703^{(6)}$ &4.91$\pm$0.29 &467.9$\pm$48.4  &1.56$\pm$0.43 &0.85$\pm$0.22 &15/13 &1524&281&1243&0.23&0.60&(23)&\\
061007(R) &$1.262^{(7)}$ &1820.0$\pm$12.0 &78.3$\pm$0.4    &2.17$\pm$0.04 &1.71$\pm$0.01 &811/79&142&31&111&0.28&0.39&(22)&\\
070318(V) &$0.84^{(8)}$  &14.20$\pm$0.43 &301.0$\pm$21.3  &1.05$\pm$0.14 &1.12$\pm$0.05 &9/6   &1090&233&857&0.27&0.77&(24)&\\
070411(R) &$2.954^{(9)}$ &3.74$\pm$2.60 &450.1$\pm$5.0   &0.76$\pm$0.03 &1.54$\pm$0.03 &754/17&1603&359&1243&0.29&0.80&(25)&\\
070419A(R)&$0.97^{(10)}$ &0.46$\pm$0.01  &587.0$\pm$20.9  &2.20$\pm$0.24 &1.27$\pm$0.04 &102/43&1387&239&1148&0.21&0.41&(26)&\\
070420(R) &...           &12.80$\pm$1.10 &213.2$\pm$18.7  &2.59$\pm$1.07 &1.10$\pm$0.33 &4/4   &433 &94&339&0.28&0.44&(23)&\\
071010A(R)&$0.98^{(11)}$ &3.90$\pm$0.26  &368.2$\pm$24.4  &2.36$\pm$0.43 &0.74$\pm$0.01 &12/20 &1234&202&1032&0.20&0.55&(27)&\\
071031(R) &$2.692^{(12)}$&0.71$\pm$0.01 &1018.6$\pm$1.6  &1.11$\pm$0.01 &0.92$\pm$0.01 &9819/22&4009&657&3352&0.20&0.65&(28)&\\
080319C(N)&$1.95^{(13)}$ &2.57$\pm$0.03 &338.3$\pm$5.6   &2.00(fixed) &1.50(fixed) &22/4  &628&137&491&0.28&0.40&(29)&\\
080330(R) &$1.51^{(14)}$ &1.43$\pm$0.02 &621.9$\pm$17.0  &0.34$\pm$0.03 &1.77$\pm$0.12 &30/36 &3552&632&2920&0.22&1.02&(30)&\\
080710(R) &$0.845^{(15)}$&3.31$\pm$0.03 &2200.9$\pm$4.1  &1.34$\pm$0.01 &0.97$\pm$0.01 &2511/61&6754&1245&5508&0.23&0.57&(31)&\\
080810(R) &$3.35^{(16)}$ &107.00$\pm$1.70 &117.6$\pm$1.1   &1.34$\pm$0.04 &1.21$\pm$0.01 &854/62&344&85&259&0.33&0.72&(32)&\\
081126(R) &...           &12.70$\pm$0.20 &201.3$\pm$1.2   &2.04$\pm$0.06 &0.44$\pm$0.01 &244/5 &1330&134&1197&0.11&0.66&(33)&\\
081203A(U)&$2.1^{(17)}$  &110.00$\pm$0.20 &367.1$\pm$0.8   &2.09$\pm$0.01 &1.49$\pm$0.01 &3176/32&794&202&592&0.34&0.55&(34)&\\

\enddata
\tablenotetext{a}{In units of $10^{-12}$erg cm$^{-2}$ s$^{-1}$.}
\tablenotetext{b}{In units of seconds.}

\tablerefs{(1) Rol et al.(2005); (2) Ledoux et al.(2005); (3) Prochaska et
al.(2006); (4) Peterson et al.(2006); (5) Ledoux et al.(2006); (6) Fugazza et
al.(2006); (7) Jakobsson et al.(2006); (8) Chen et al.(2007); (9) Jakobsson et
al.(2007); (10) Cenko et al.(2007); (11) Prochaska et al.(2007); (12) Ledoux et
al.(2007); (13) Wiersema et al.(2008); (14) Cucchiara (2008); (15) Perley et
al.(2008); (16) Prochaska et al.(2008); (17) Landsman et al.(2008); (18) Rykoff
et al.(2004); (19) Pandey et al.(2006); (20) Cenko et al.(2007); (21) Molinari
et al.(2007); (22) Rykoff et al.(2009); (23) Klotz et al.(2008); (24) Roming et
al.(2009); (25) Malesani  et al.(2007); (26) Melandri et al.(2009); (27) Covino
et al.(2008); (28) Kr{\"u}hler et al.(2009); (29) Holland  et al.(2008); (30)
Guidorzi et al.(2009); (31) Kr{\"u}hler et al.(2009); (32) Page et al.(2009);
(33) Klotz et al.(2009); (34) Kuin et al.(2009) }

\end{deluxetable}

\begin{deluxetable}{lccccccccccccccccccccccc}

\tablewidth{450pt} \tabletypesize{\tiny}
\tablecaption{XRT observations and fitting result for our sample} \tablenum{2}

\tablehead{\colhead{GRB}&\colhead{$z^{Ref}$}&\colhead{$F_m$
\tablenotemark{a}}&\colhead{$t_{p}$\tablenotemark{b}}&\colhead{$\alpha_1$}&
\colhead{$\alpha_2$}&\colhead{$\chi^2$(dof)}
&\colhead{$\omega$}&\colhead{$t_r$\tablenotemark{b}}&\colhead{$t_d$\tablenotemark{b}}&\colhead{$t_r/t_d$}&\colhead{$t_r/t_p$}}

\startdata
060319 &...           &8.46$\pm$0.98 &267.0$\pm$20.2  &5.46$\pm$2.05 &1.13$\pm$0.04 &43/30  &490&82&408&0.20&0.31&\\
060801 &...           &20.20$\pm$3.50 &114.3$\pm$9.5   &6.69$\pm$4.42 &1.77$\pm$0.27 &3/6    &134&39&94&0.42&0.34&\\
060804 &...           &7.60$\pm$0.66 &418.9$\pm$176.1 &0.85$\pm$0.67 &1.22$\pm$0.07 &25/23  &1603&281&1322&0.21&0.67&\\
070103 &...           &1.60$\pm$0.13 &685.5$\pm$64.3  &0.76$\pm$0.16 &1.47$\pm$0.08 &22/24  &2325&521&1803&0.29&0.76&\\
070208 &$1.165^{(1)}$ &1.80$\pm$0.29 &968.1$\pm$72.9  &1.09$\pm$0.27 &1.29$\pm$0.06 &39/30  &3274&657&2616&0.25&0.68&\\
070714A&...           &1.38$\pm$0.24 &234.0$\pm$36.9  &2.25$\pm$1.46 &0.86$\pm$0.12 &16/8   &685&94&591&0.16&0.40&\\
080307 &...           &67.10$\pm$1.50 &210.5$\pm$3.12  &2.22$\pm$0.16 &2.05$\pm$0.04 &141/153&338&94&244&0.39&0.45&\\
080319C&$1.95^{(2)}$  &58.40$\pm$3.50 &432.8$\pm$29.1  &1.55$\pm$0.41 &1.41$\pm$0.03 &72/52  &994&281&713&0.39&0.65&\\
080409 &...           &0.78$\pm$0.08 &395.9$\pm$100.0 &0.50$\pm$0.00 &1.11$\pm$0.11 &13/10  &2007&336&1671&0.20&0.85&\\
090429B&...           &1.52$\pm$0.14 &540.2$\pm$51.6  &1.57$\pm$0.38 &1.34$\pm$0.08 &15/13  &1487 &339&1148&0.29&0.63&\\
090607 &...           &40.40$\pm$2.50 &118.9$\pm$3.48  &4.86$\pm$1.13 &2.79$\pm$0.39 &25/20  &134&39&94&0.42&0.33&\\
\enddata
\tablenotetext{a}{In units of $10^{-11}$erg cm$^{-2}$ s$^{-1}$.}
\tablenotetext{b}{In units of seconds.} \tablerefs{(1) Cucchiara et al.(2007);
(2) Prochaska et al.(2007); (3) Wiersema et al.(2008) }

\end{deluxetable}


\begin{deluxetable}{lccccccccccccc}
\tablewidth{450pt} 
\tablecaption{The optical and X-ray with know redshift in our sample}
\tablenum{3}

\tablehead{\colhead{GRB}&\colhead{$z$}&\colhead{$E_{\rm
iso,\gamma}$}&\colhead{$L_{\rm p}$}&\colhead{$E_{\rm
iso,O}$}&\colhead{$\Gamma_0$}& \colhead{$R_{\rm
d}$}\\\colhead{}&\colhead{}&\colhead{($10^{52}$ erg)}&\colhead{($10^{47}$
erg/s)}&\colhead{($10^{48}$ erg)}&\colhead{}& \colhead{($10^{17}$cm)}}

\startdata
Optical&&&&\\
\hline
050730  &3.97  &13.37$\pm$1.22 &166.42&6.04$\pm$0.68  &306$\pm$26 &1.67\\
050820A &2.615 &2.92$\pm$0.31  &170.74&11.72$\pm$0.69 &262$\pm$5  &1.11\\
060418  &1.49  &4.68$\pm$0.14  &111.23&12.42$\pm$0.24 &344$\pm$3  &1.09\\
060605  &3.8   &1.98$\pm$0.26  &274.19&13.05$\pm$0.27 &276$\pm$6  &0.95\\
060607A &3.082 &5.25$\pm$0.20  &114.90&23.56$\pm$0.40 &394$\pm$3  &1.03\\
060904B &0.703 &0.21$\pm$0.02  &7.18&0.11$\pm$0.01  &133$\pm$5  &0.73\\
061007  &1.262 &75.96$\pm$4.93 &800.64&168.6$\pm$1.12 &603$\pm$5  &1.89\\
070318  &0.84  &0.46$\pm$0.02  &20.78&0.48$\pm$0.01  &178$\pm$5  &0.78\\
070411  &2.954 &5.09$\pm$0.30  &83.64&2.78$\pm$0.02  &276$\pm$2  &1.3\\
070419A &0.97  &0.14$\pm$0.02  &1.13&0.02$\pm$0.006 &122$\pm$3  &0.67\\
071010A &0.98  &0.05$\pm$0.01  &8.67&0.19$\pm$0.01  &129$\pm$5  &0.46\\
071031  &2.692 &1.45$\pm$0.21  &33.26&0.42$\pm$0.03  &169$\pm$3  &1.18\\
080319C &1.95  &13.8$\pm$3.13  &11.55&0.70$\pm$0.01  &311$\pm$9  &1.67\\
080330  &1.51  &0.20$\pm$0.05  &21.24&0.21$\pm$0.003 &137$\pm$4  &0.70\\
080710  &0.845 &0.26$\pm$0.04  &31.70&0.11$\pm$0.001 &79$\pm$1   &1.11\\
080810  &3.35  &39.37$\pm$2.32 &629.20&107.6$\pm$1.71 &610$\pm$5  &1.51\\
081203A &2.1   &8.08$\pm$0.31  &666.19&35.66$\pm$0.08 &288$\pm$1  &1.47\\

\hline
X-ray&&&&\\
\hline
070208  &1.165 &0.16$\pm$0.04&1.40$\pm$0.12 &194.6 &102$\pm$3    &0.79\\
080319C &1.95  &3.31$\pm$0.09  &162$\pm$8&3772.5 &242$\pm$4    &1.23
\enddata
\end{deluxetable}


\begin{deluxetable}{lccccccccccccc}
\tablewidth{350pt} 
\label{correlations}
\tablecaption{Spearman correlation coefficients of the Optical Selected Sample}
\tablenum{4}

\tablehead{ \colhead{}& \colhead{$L_{\rm p,O}$}& \colhead{${t_p}^{'}$}&
\colhead{$r$}& \colhead{$d$}& \colhead{${t_{r}}^{'}$} &
\colhead{${t_{d}}^{'}$}& \colhead{$t_r/t_d$}& \colhead{$t_r/t_p$}&
\colhead{$w^{'}$}} \startdata
$L_{\rm p,O}$       & &-0.90&X&X&-0.89&-0.88&X   &X   &-0.88\\
${t_p}^{'}$     & &     &X&X&0.95 &0.93 &X   &X   &0.94\\
$r$             & &     & &X&X    &X    &0.90&0.90&X   \\
$d$             & &     & & &X    &X    &X   &X    &X\\
${t_r}^{'}$     && &&&&0.98&X&X&0.98\\
${t_d}^{'}$     & &&&&&&X&X&$\sim 1$\\
$t_r/t_d$       & & &&&&&&X&X\\
$t_r/t_p$       & & &&&&&&&X\\
\enddata
\end{deluxetable}

\clearpage \thispagestyle{empty} \setlength{\voffset}{-18mm}

\begin{figure*}
\includegraphics[angle=0,scale=0.350]{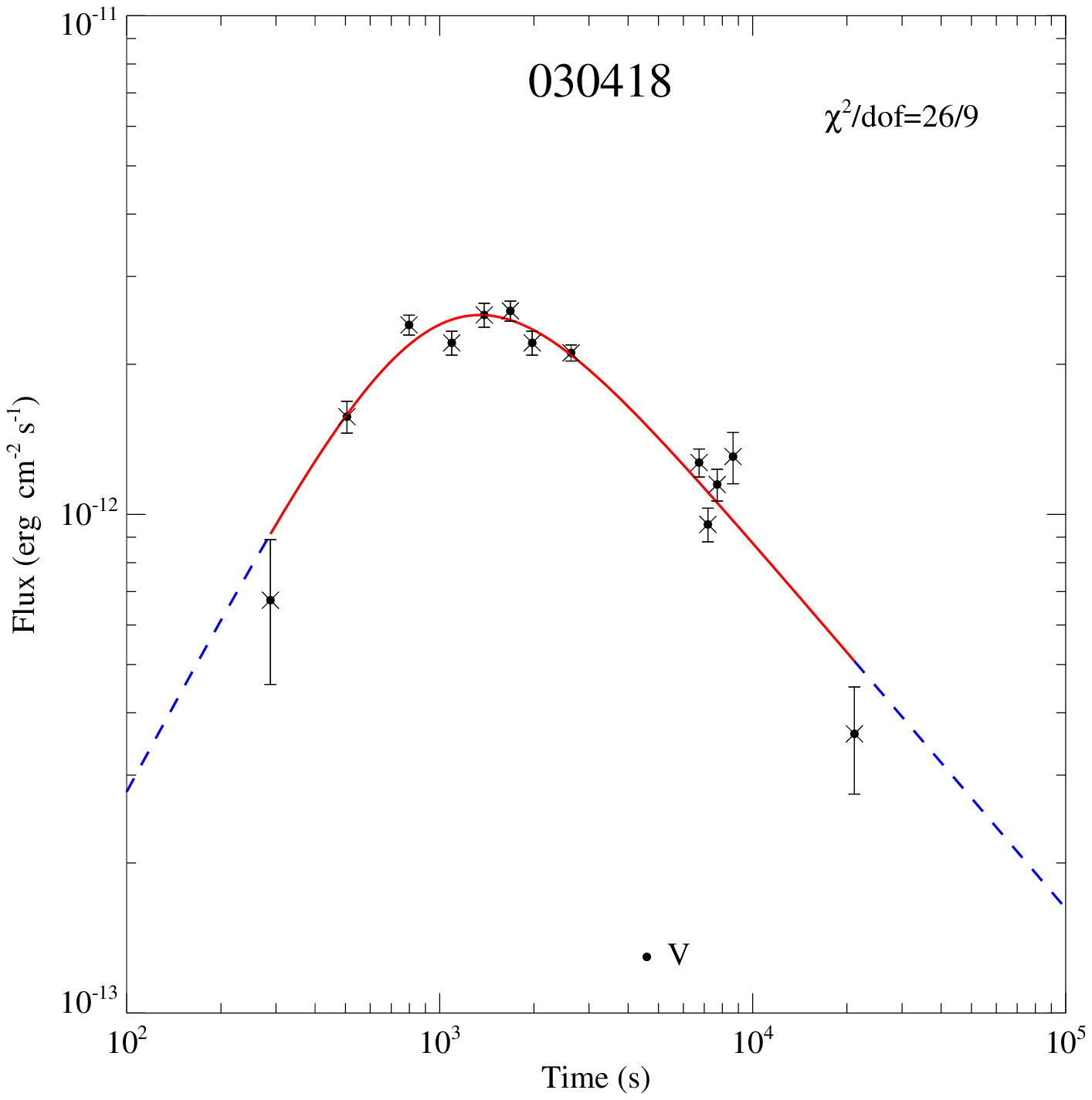}
\includegraphics[angle=0,scale=0.350]{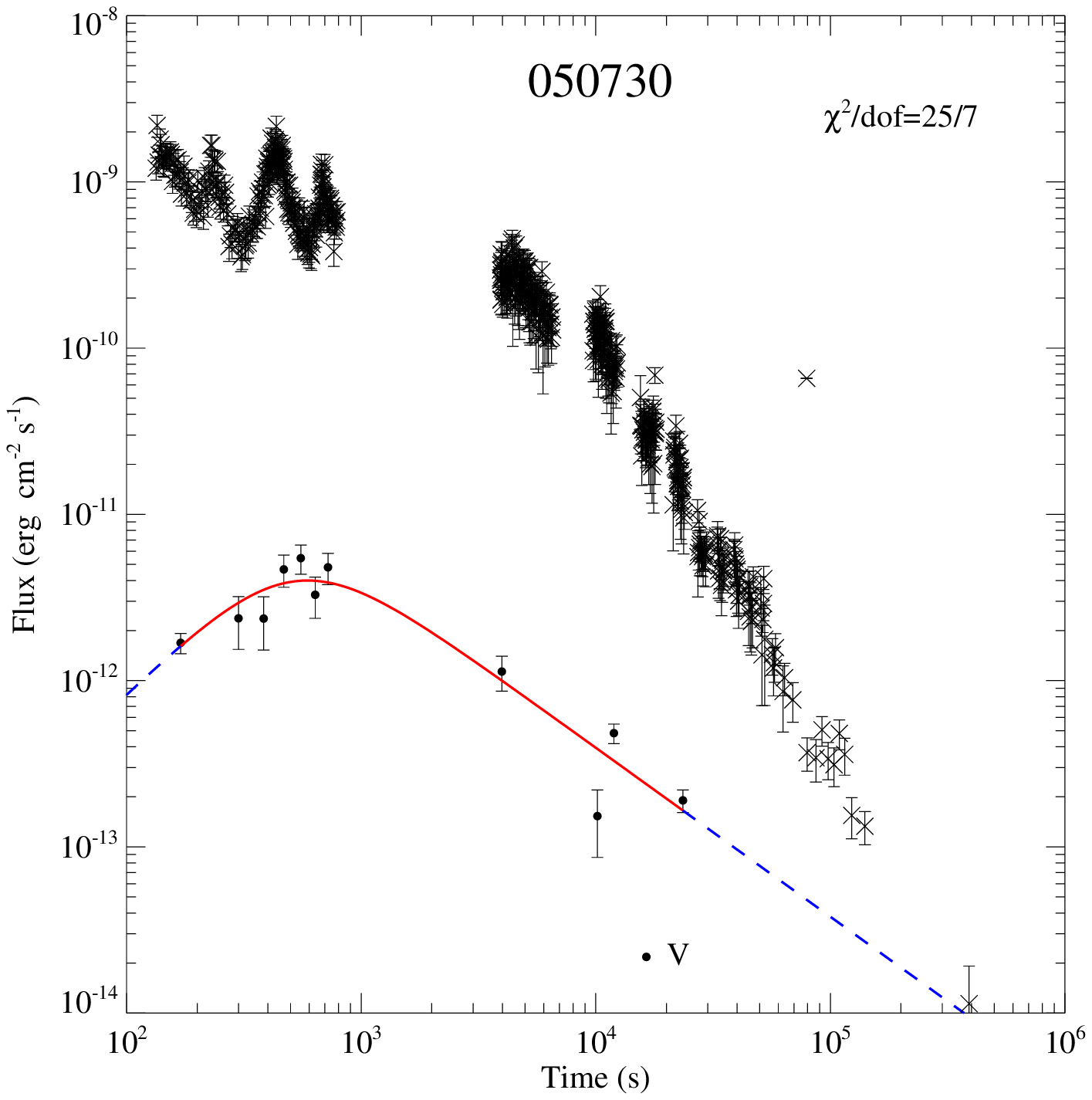}
\includegraphics[angle=0,scale=0.350]{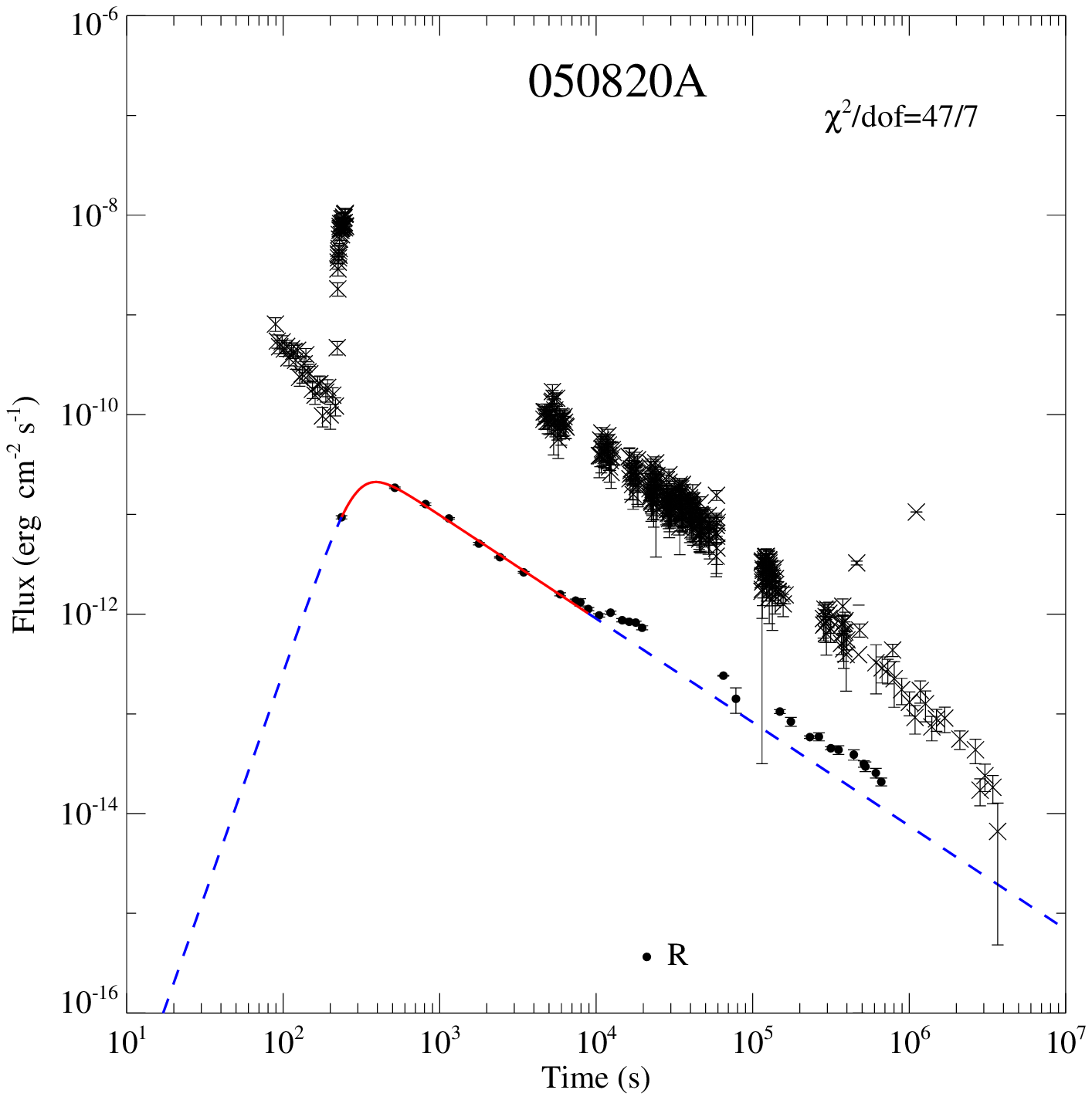}
\includegraphics[angle=0,scale=0.350]{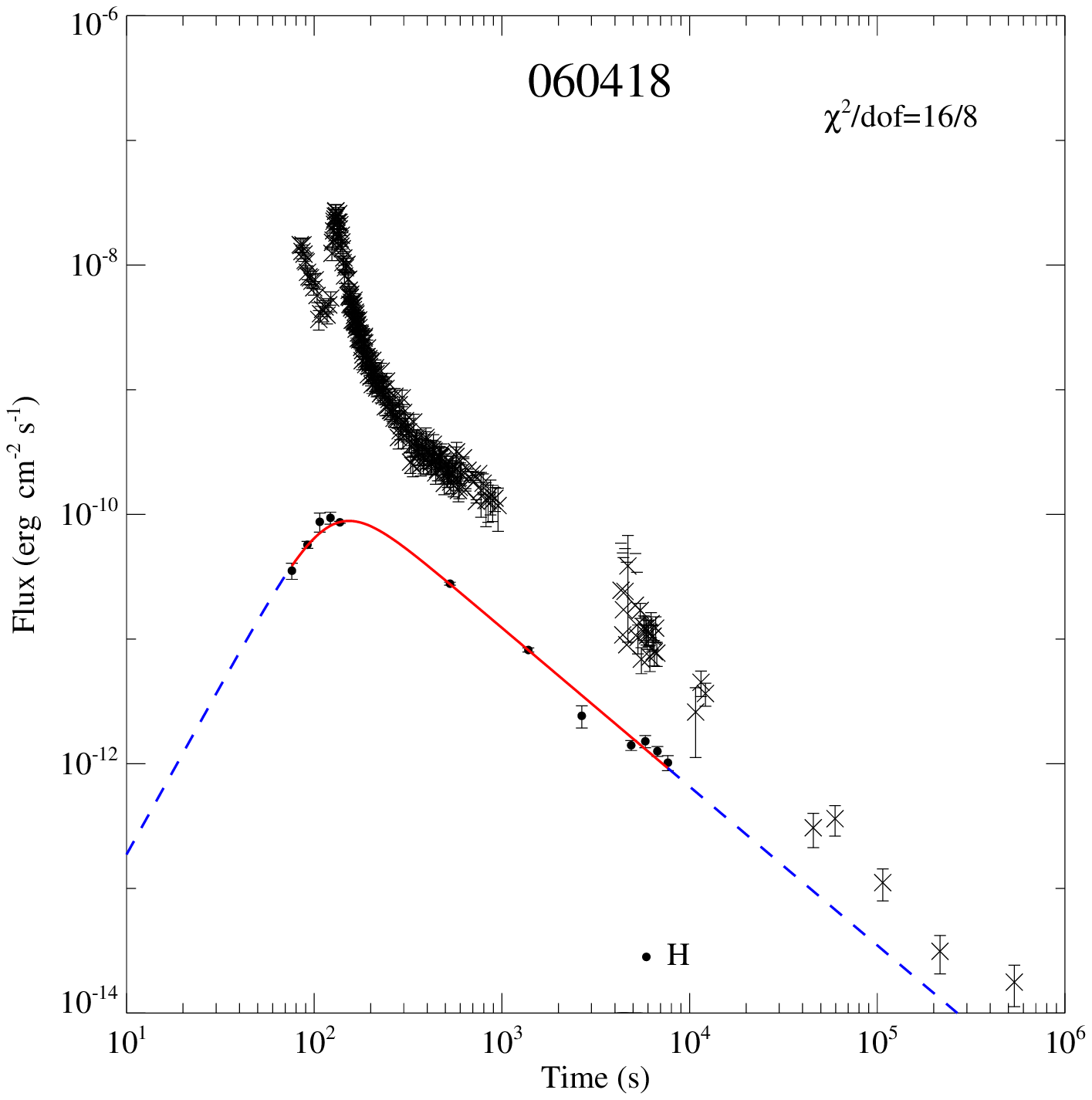}
\includegraphics[angle=0,scale=0.350]{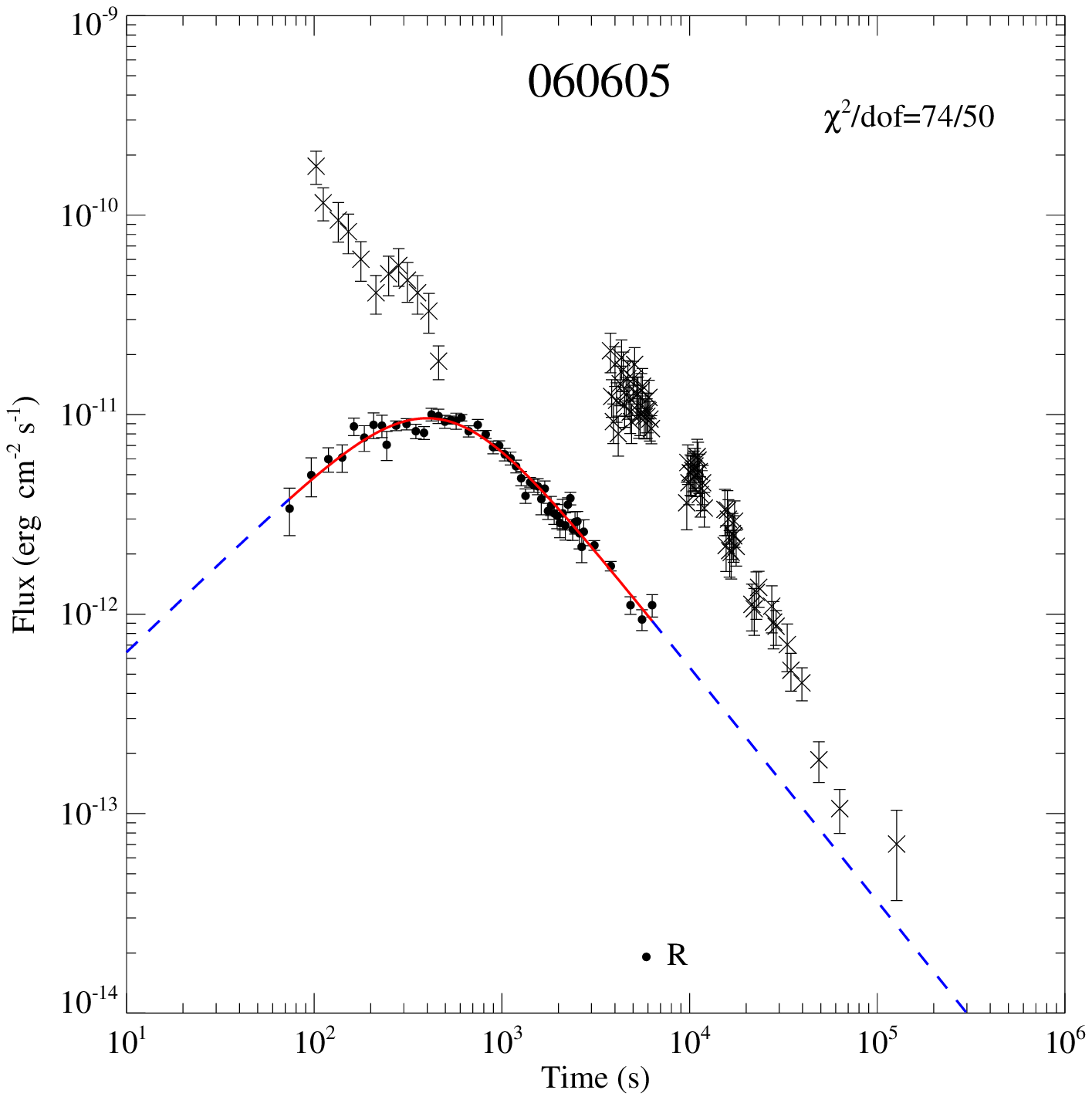}
\includegraphics[angle=0,scale=0.350]{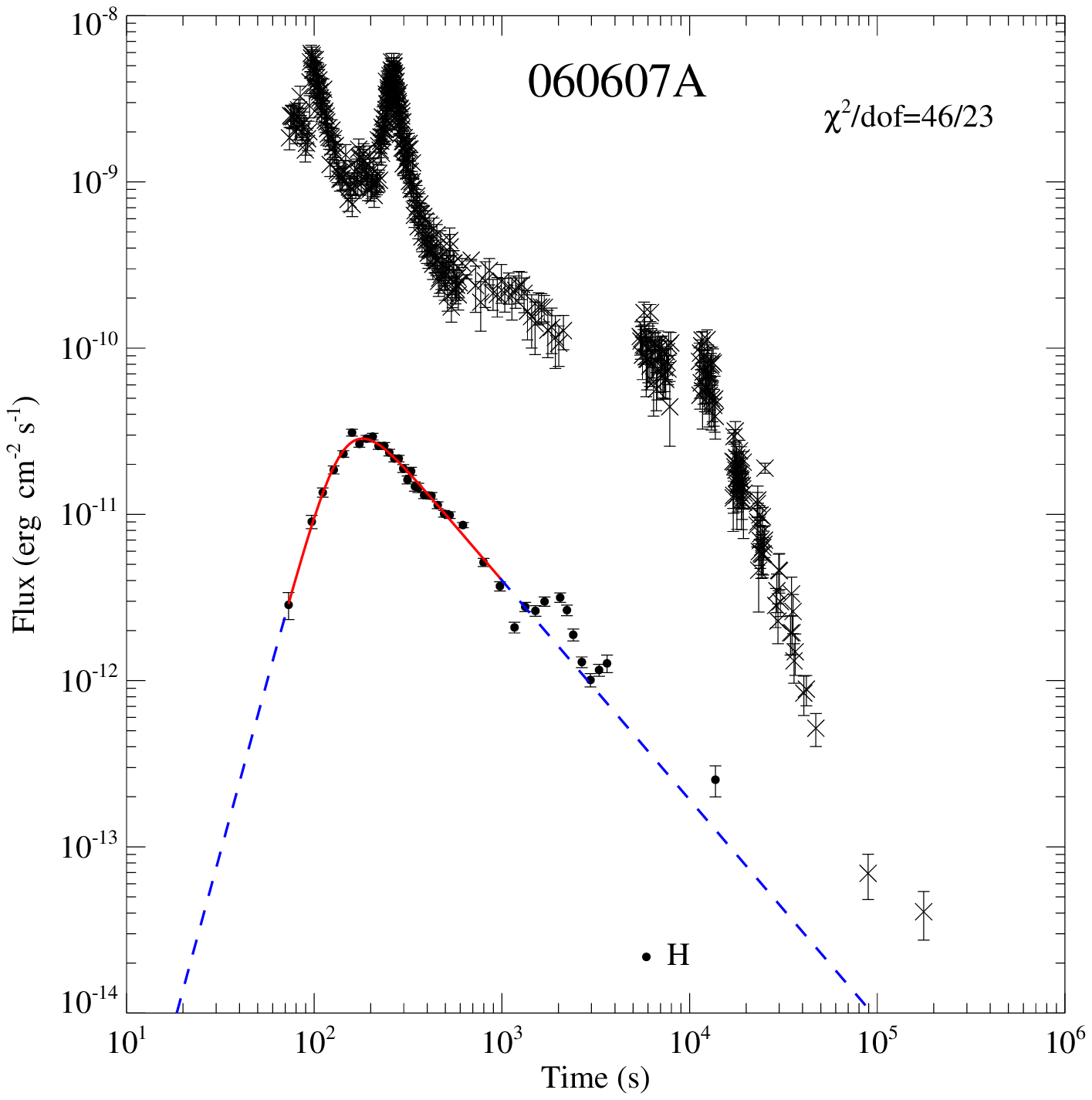}
\includegraphics[angle=0,scale=0.350]{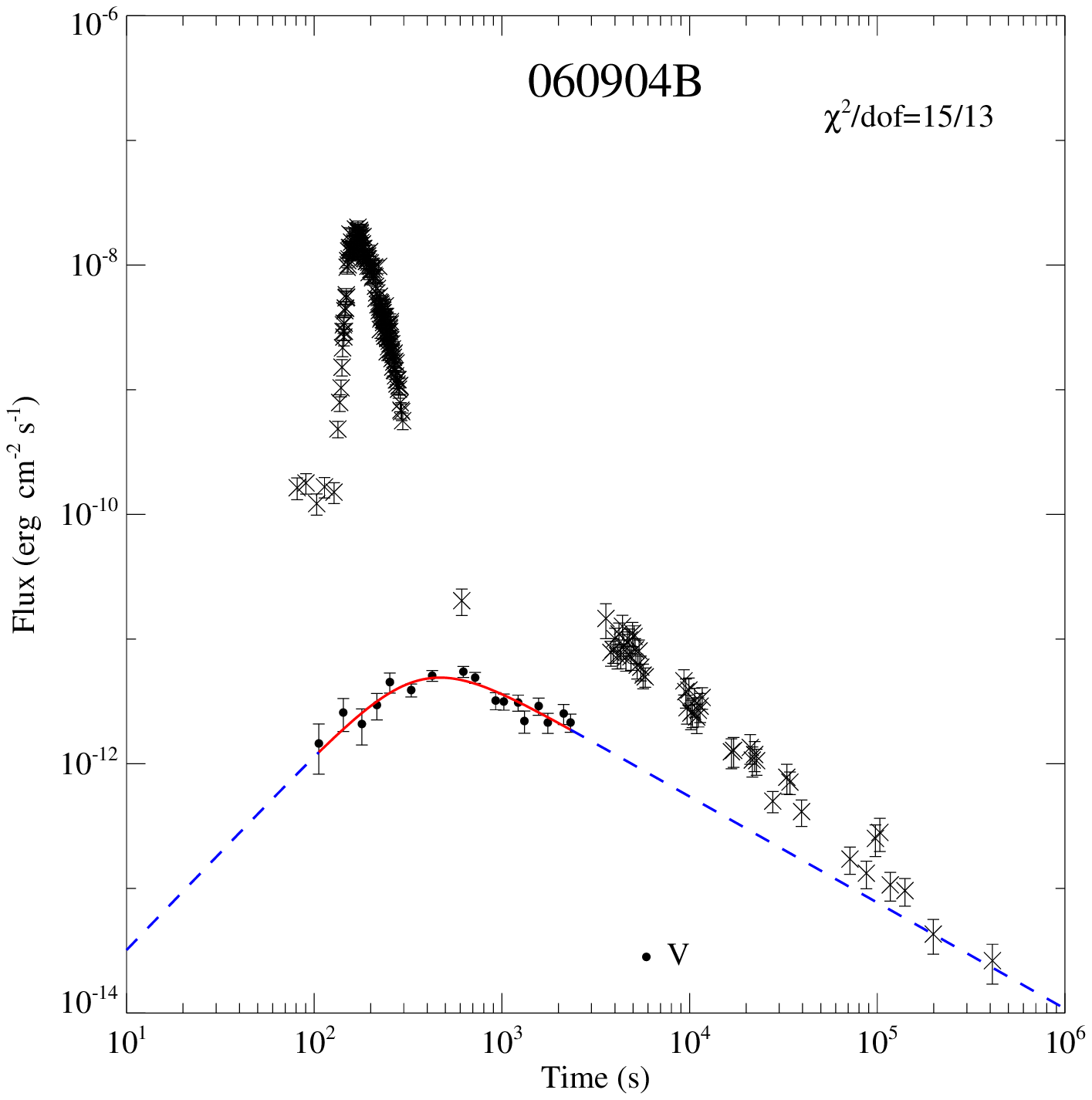}
\includegraphics[angle=0,scale=0.350]{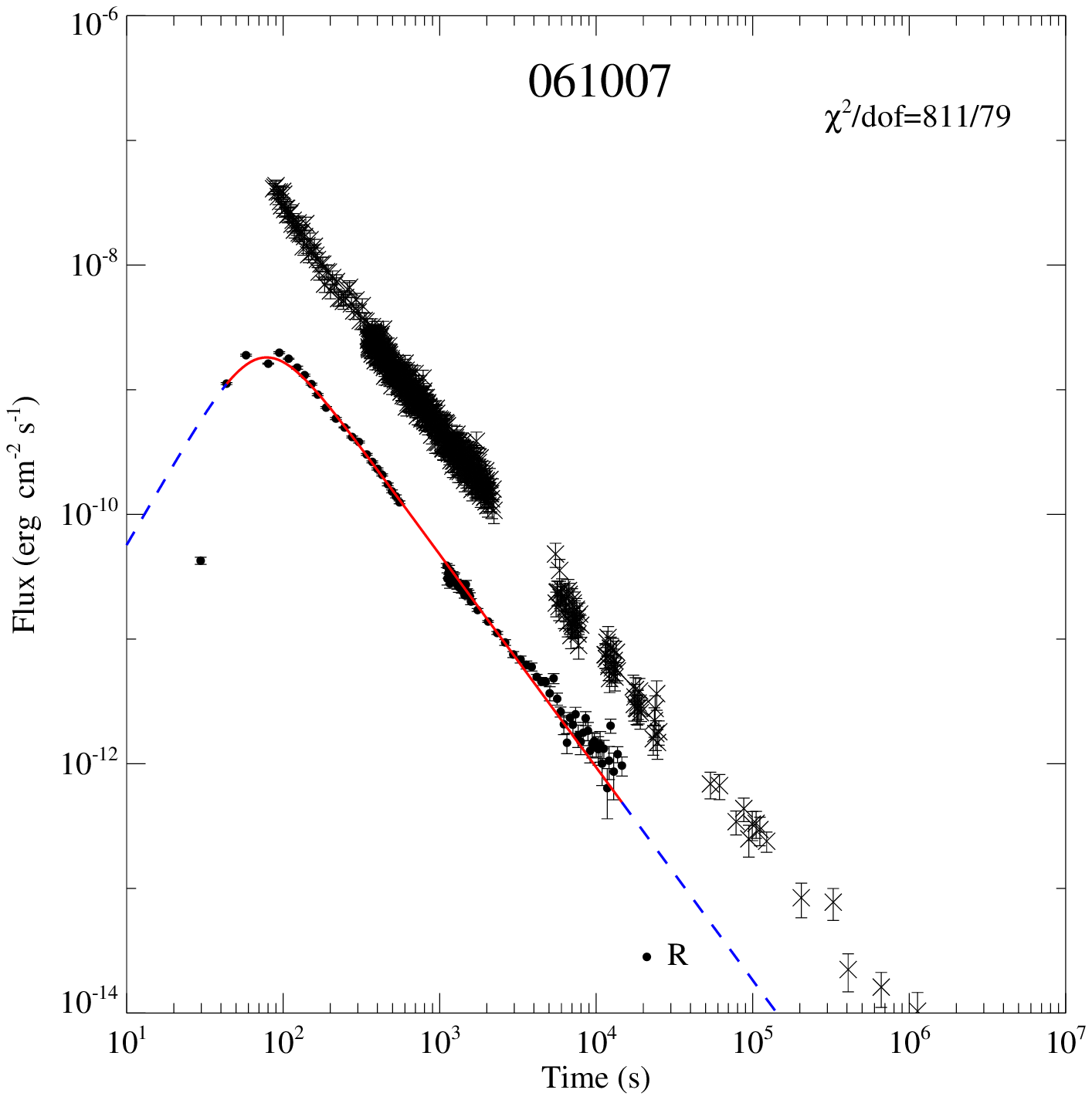}
\includegraphics[angle=0,scale=0.350]{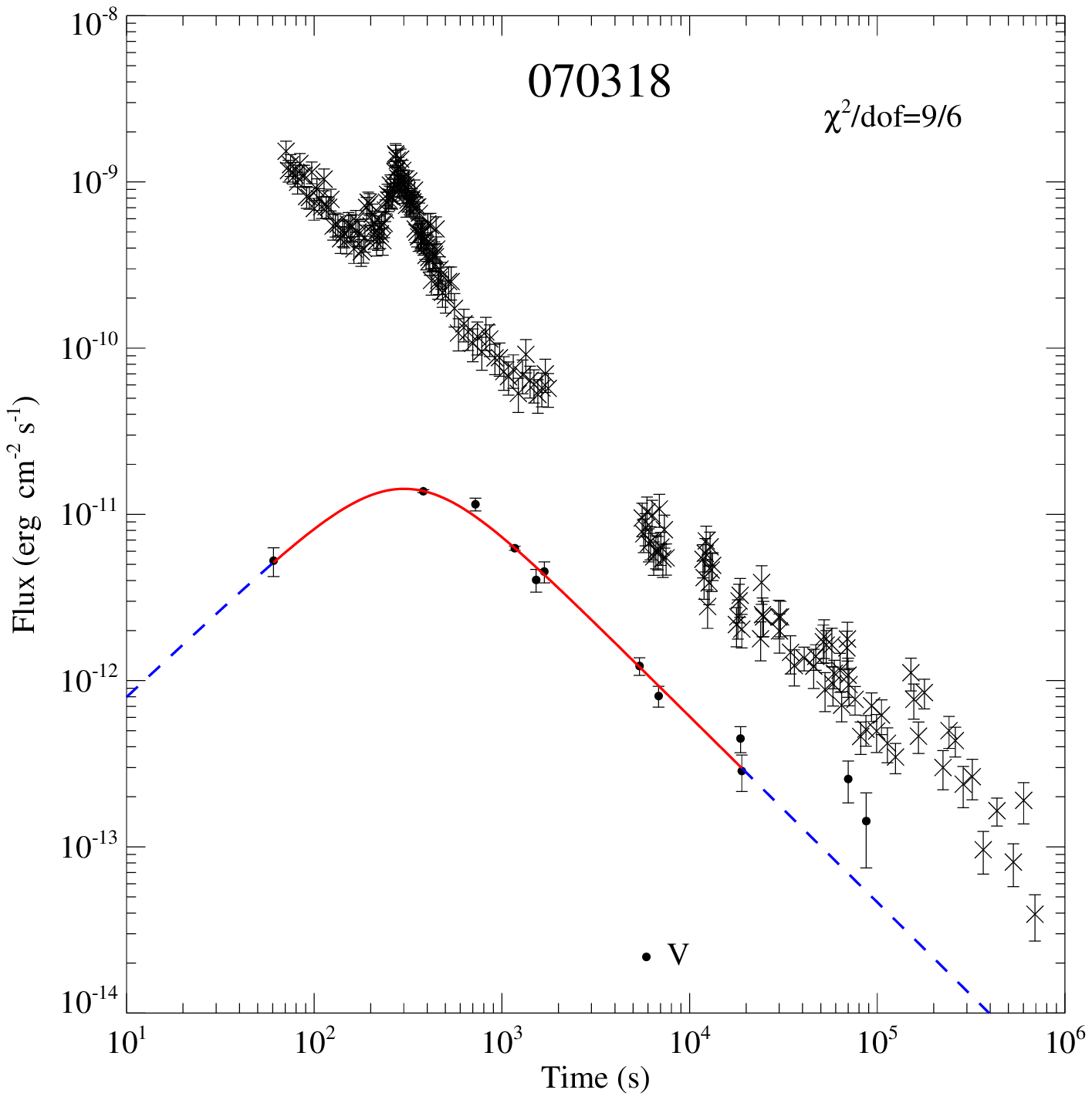}
\includegraphics[angle=0,scale=0.350]{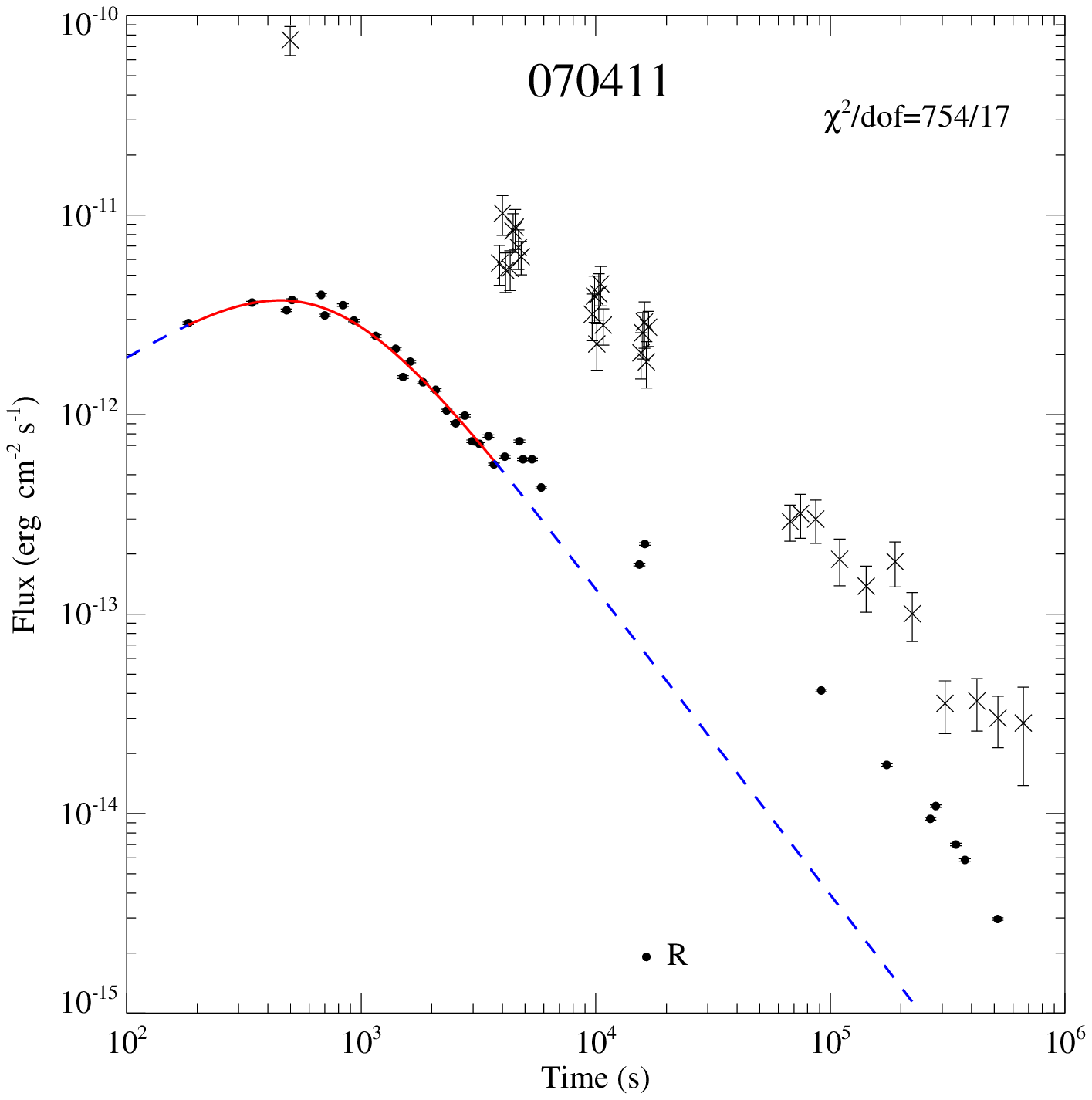}
\includegraphics[angle=0,scale=0.350]{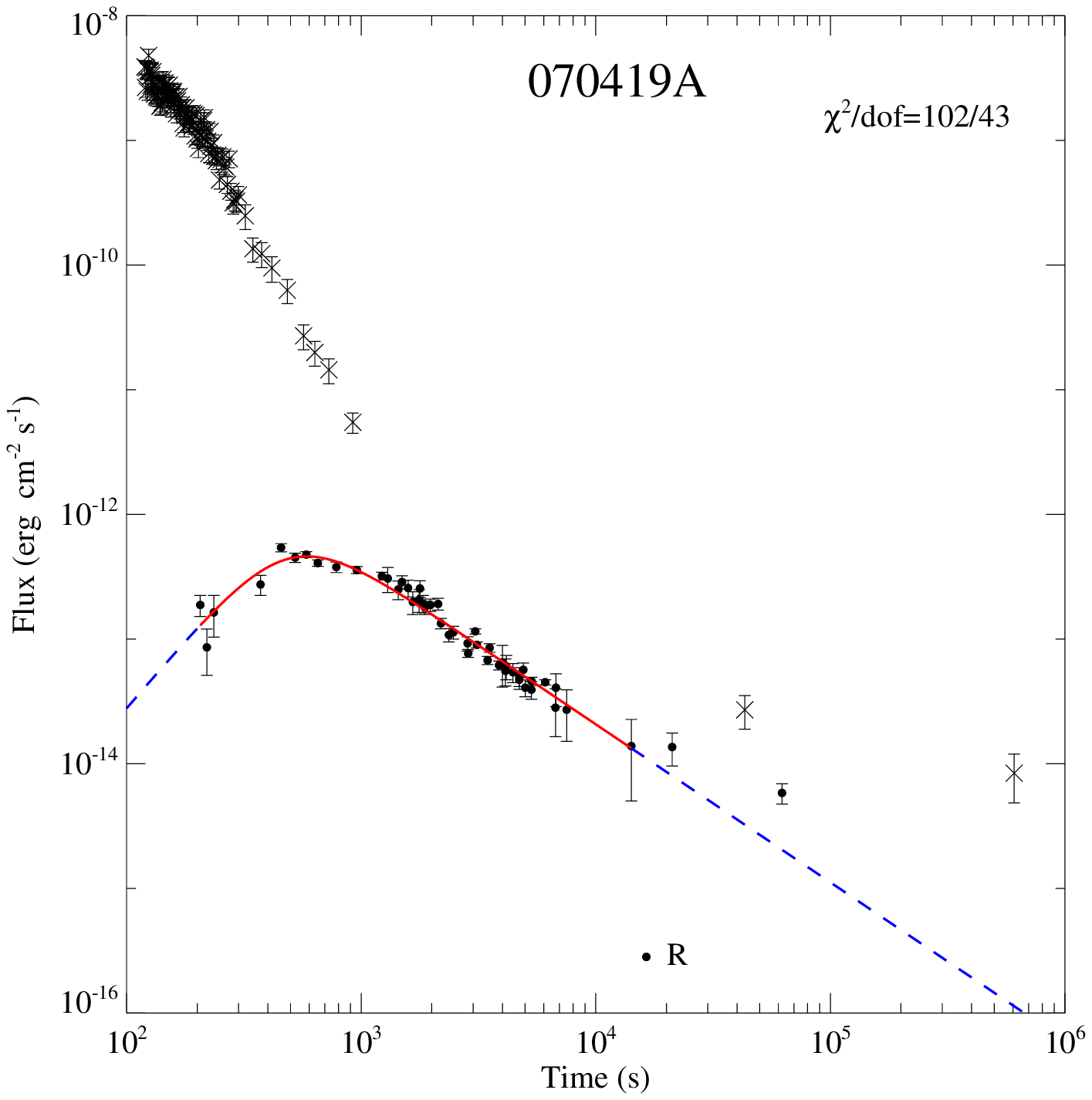}
\hfill
\includegraphics[angle=0,scale=0.350]{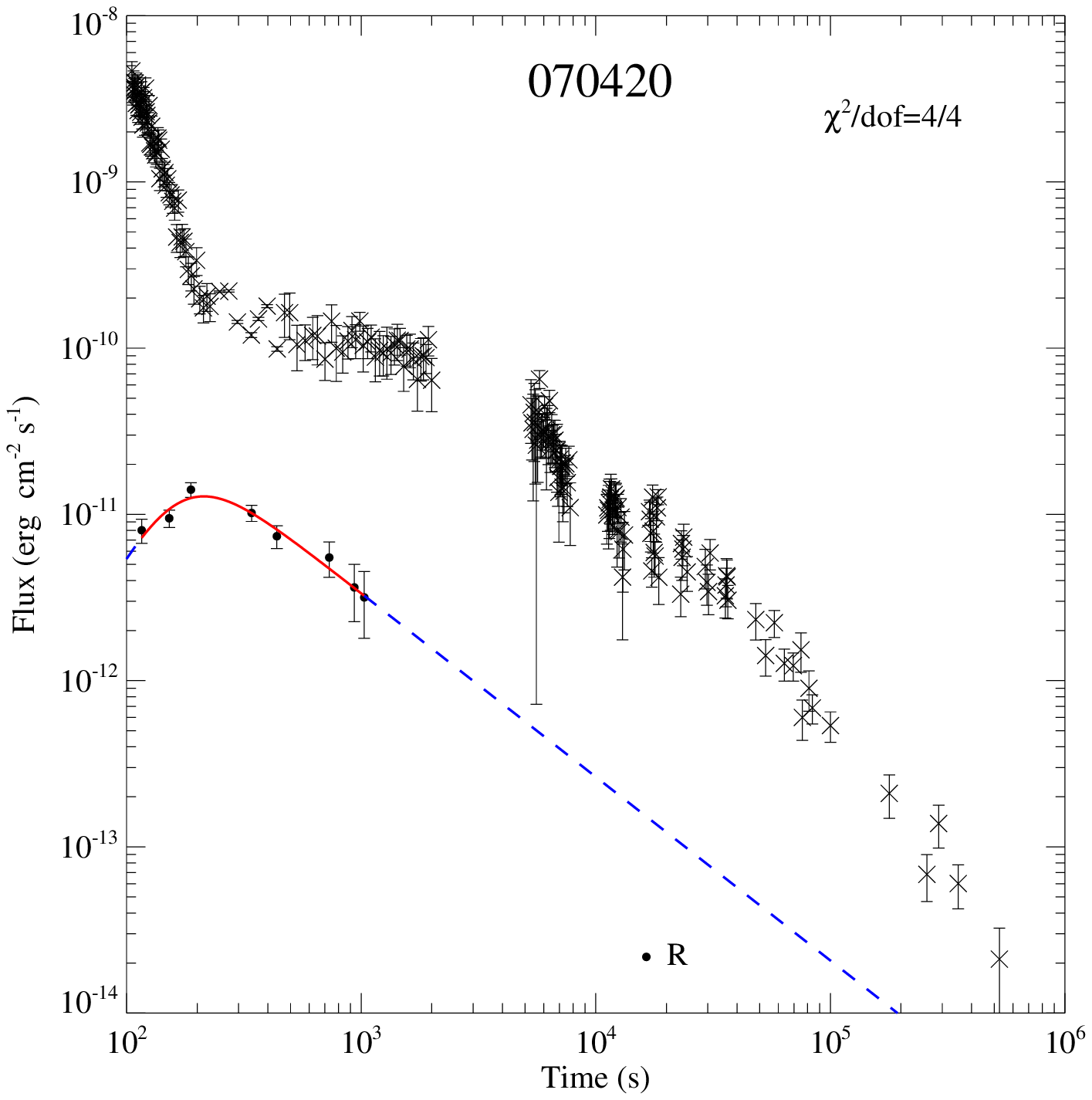}

\caption{Lightcurves of the optical-selected sample (dots). Lines are the best
fits  with a smooth broken power-law to the optical data in a selected interval
marked as a red {\em solid} segment. Blue {\em dashed} segments just extend the
fitting curves in a broad time regime. The simultaneous X-ray data observed
with {\em Swift}/XRT (crosses with error bands) are also present.}
\label{optical}
\end{figure*}
\clearpage \setlength{\voffset}{0mm}
%

\begin{figure*}
\includegraphics[angle=0,scale=0.350]{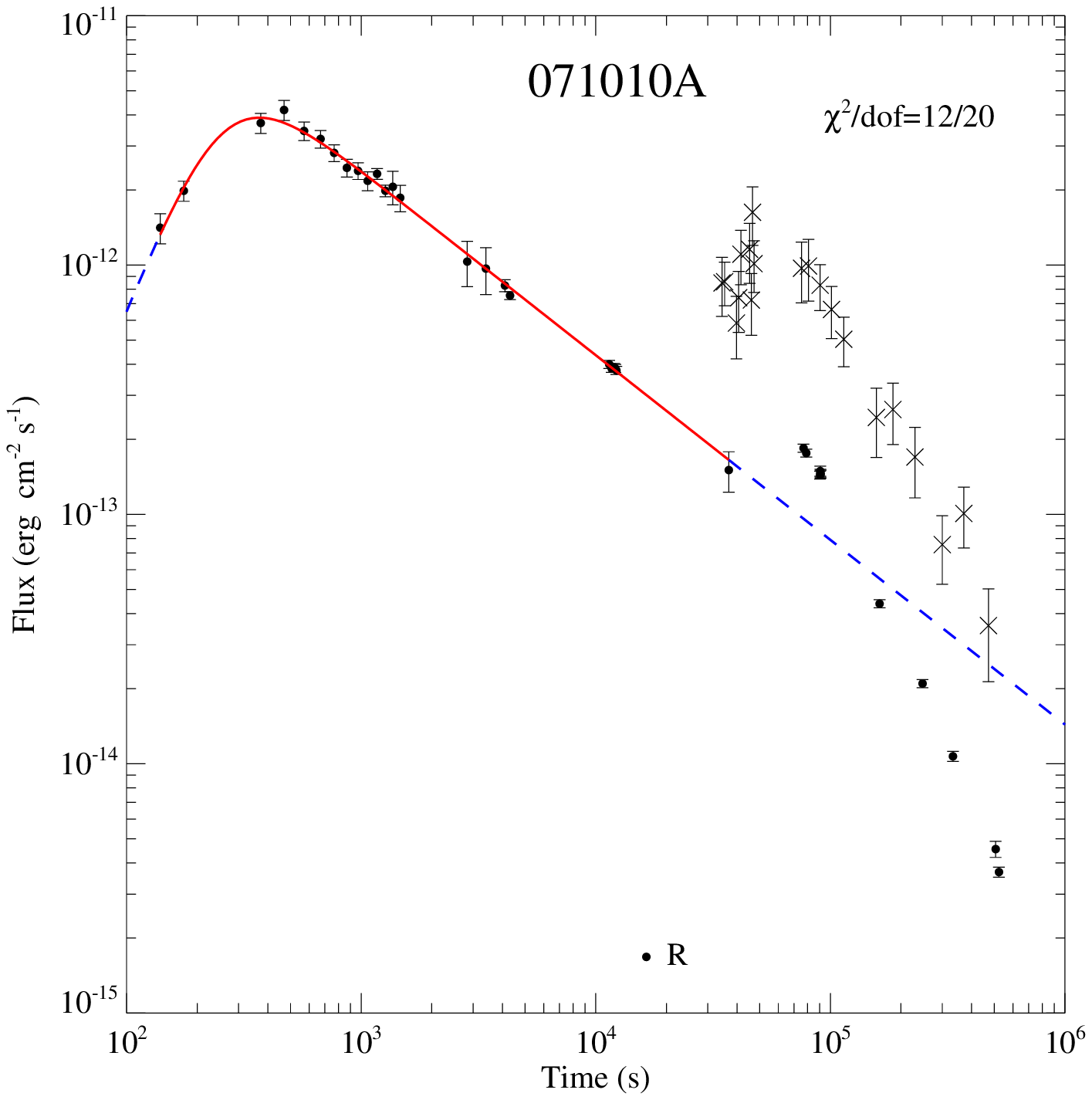}
\includegraphics[angle=0,scale=0.350]{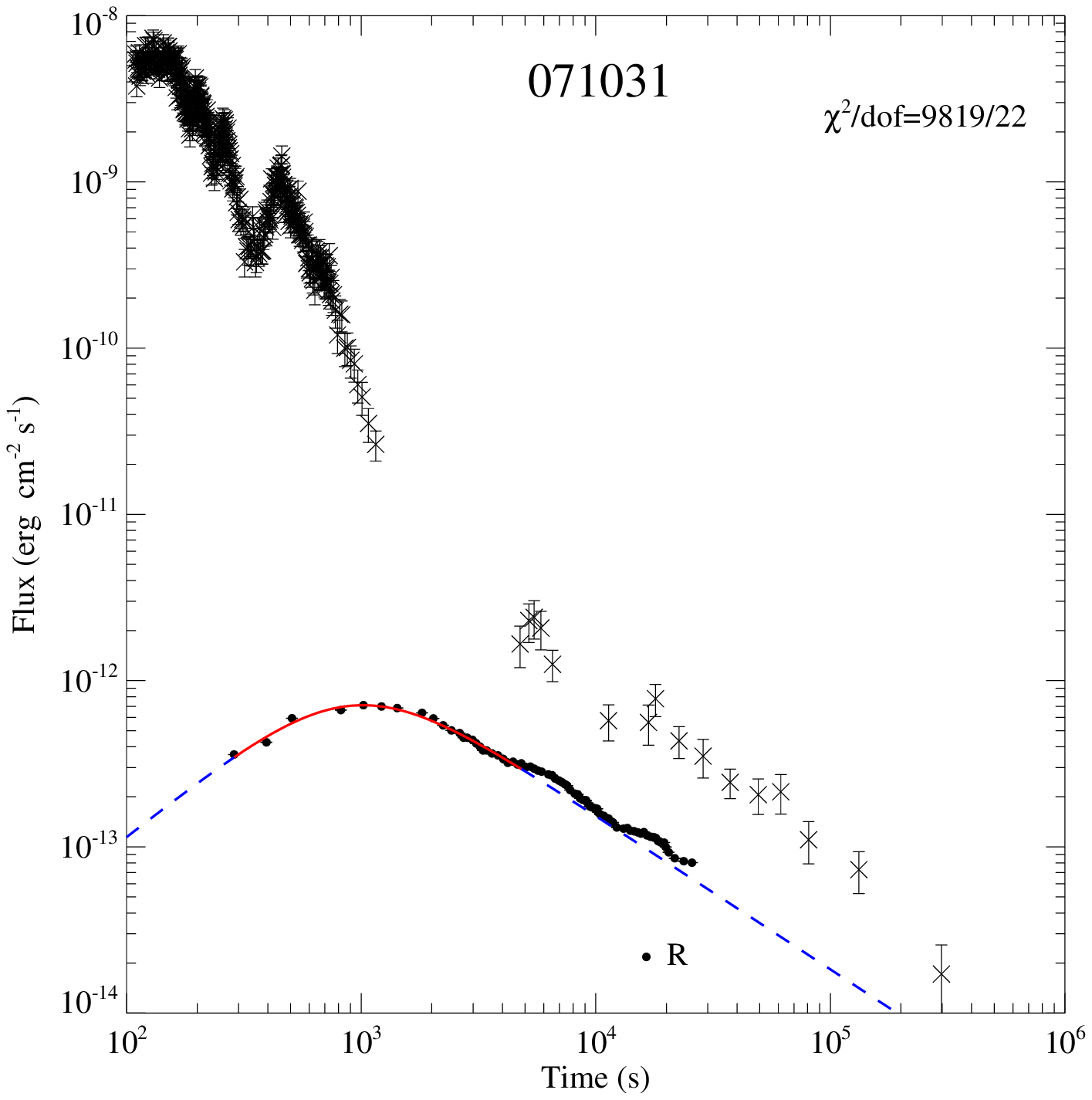}
\includegraphics[angle=0,scale=0.350]{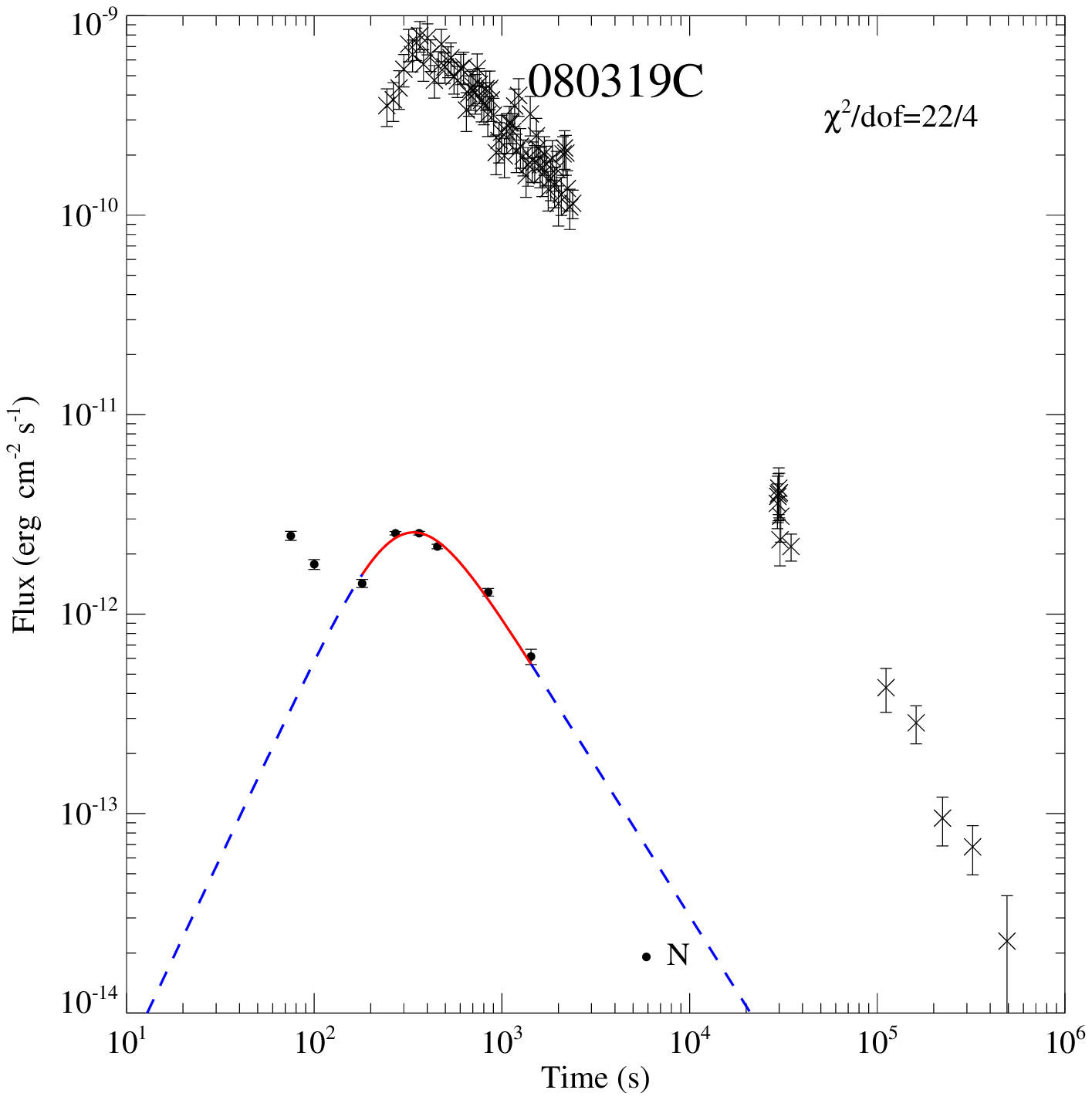}
\includegraphics[angle=0,scale=0.350]{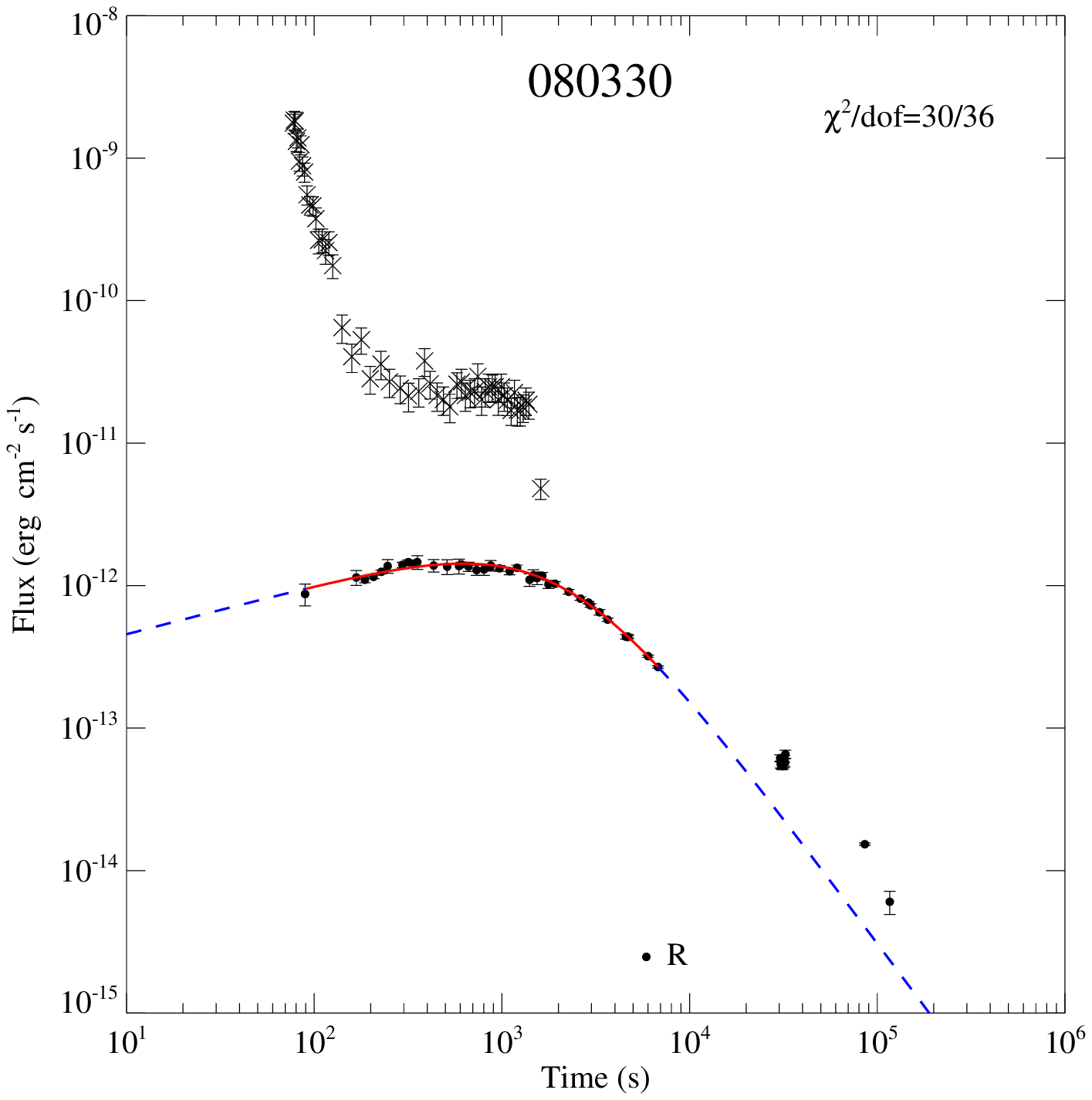}
\includegraphics[angle=0,scale=0.350]{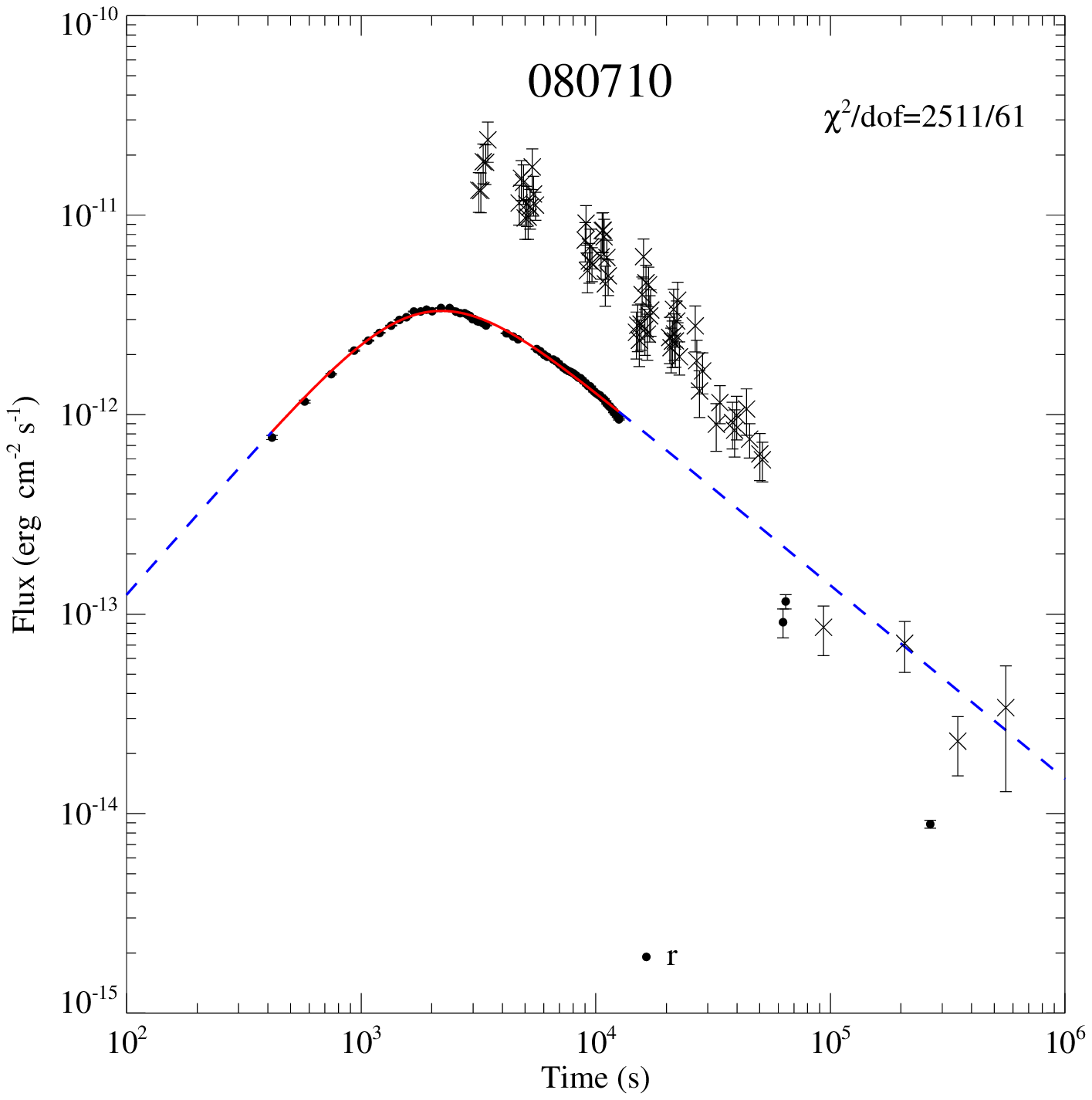}
\includegraphics[angle=0,scale=0.350]{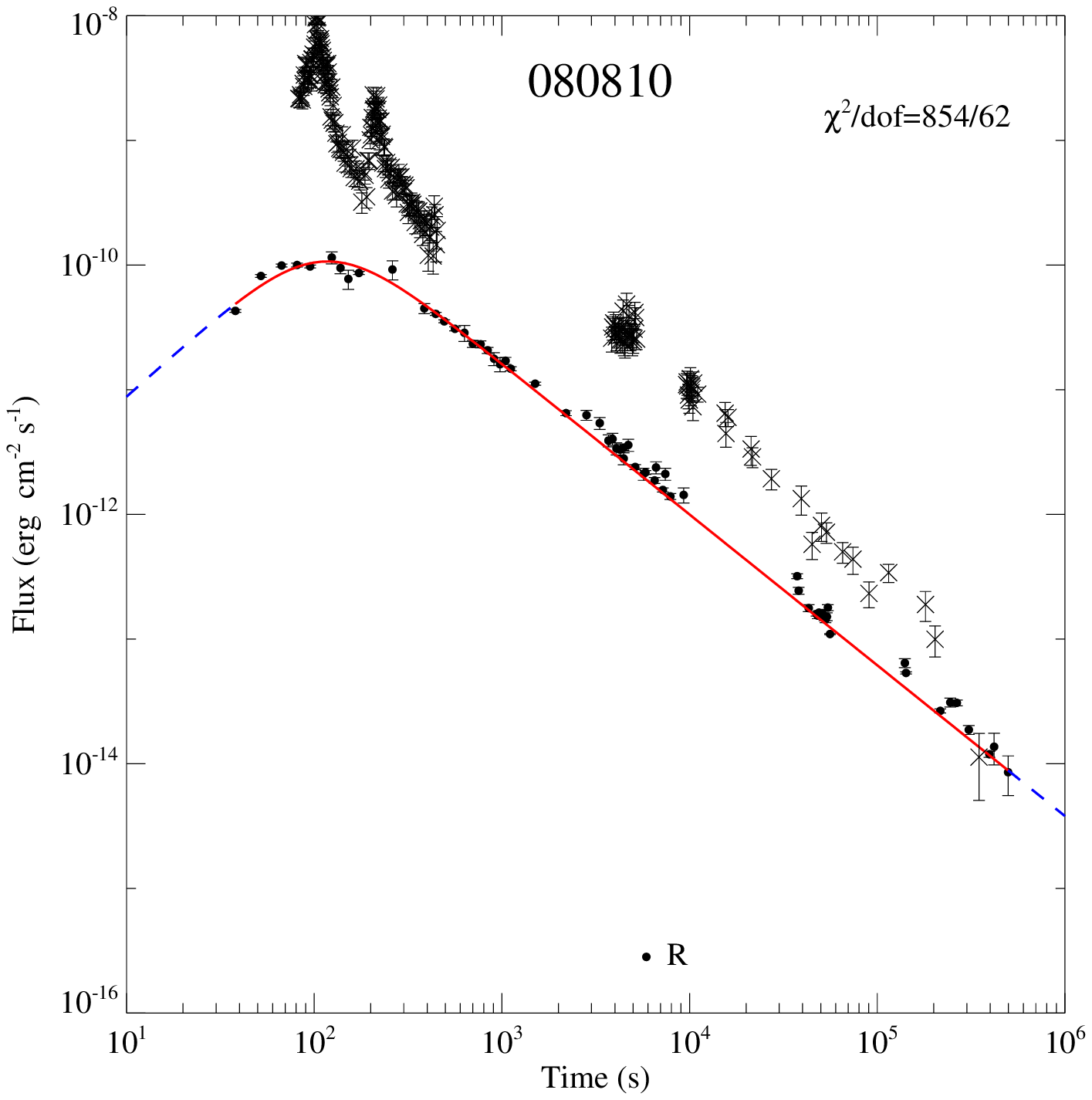}
\includegraphics[angle=0,scale=0.350]{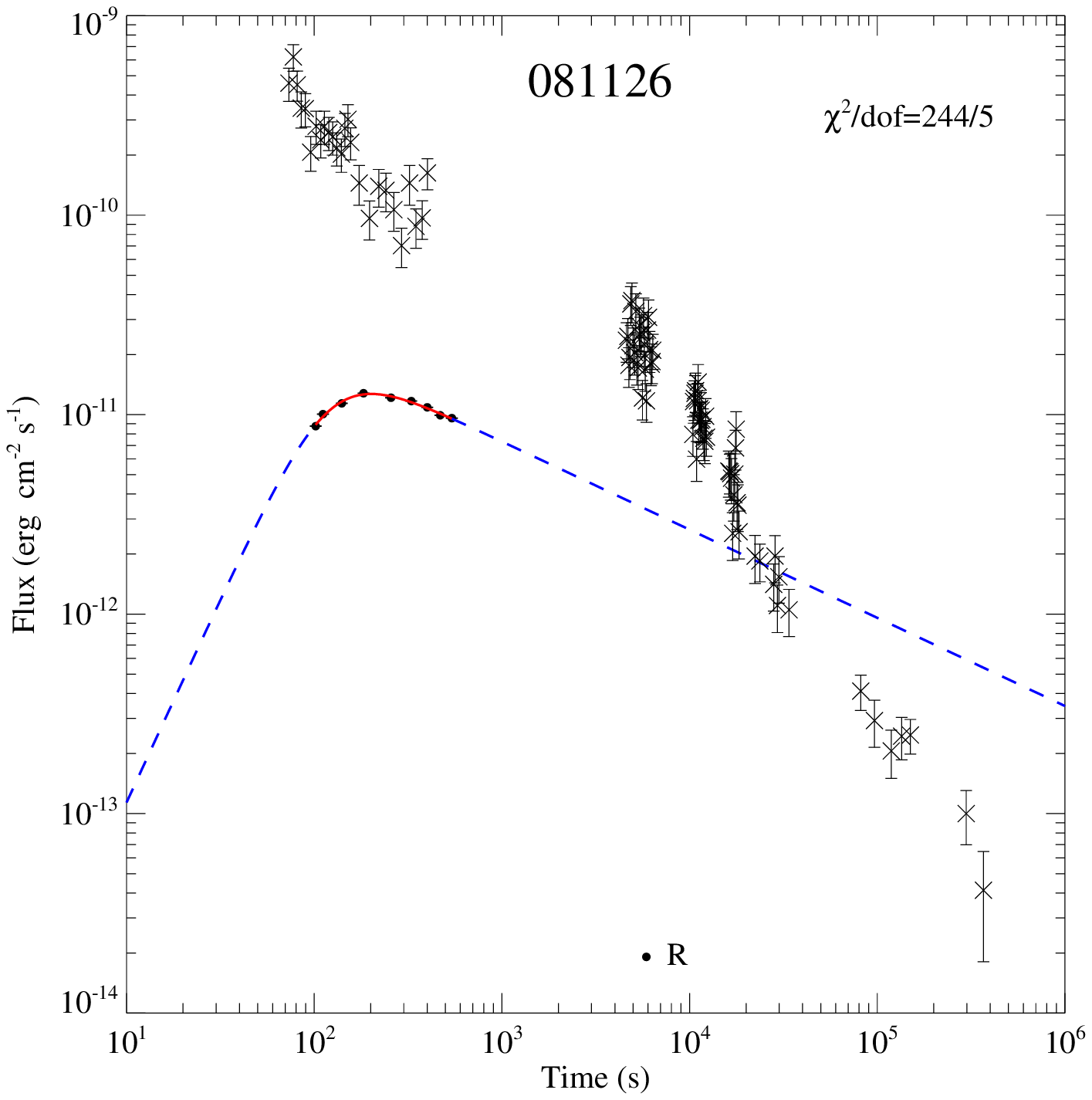}
\hfill
\includegraphics[angle=0,scale=0.350]{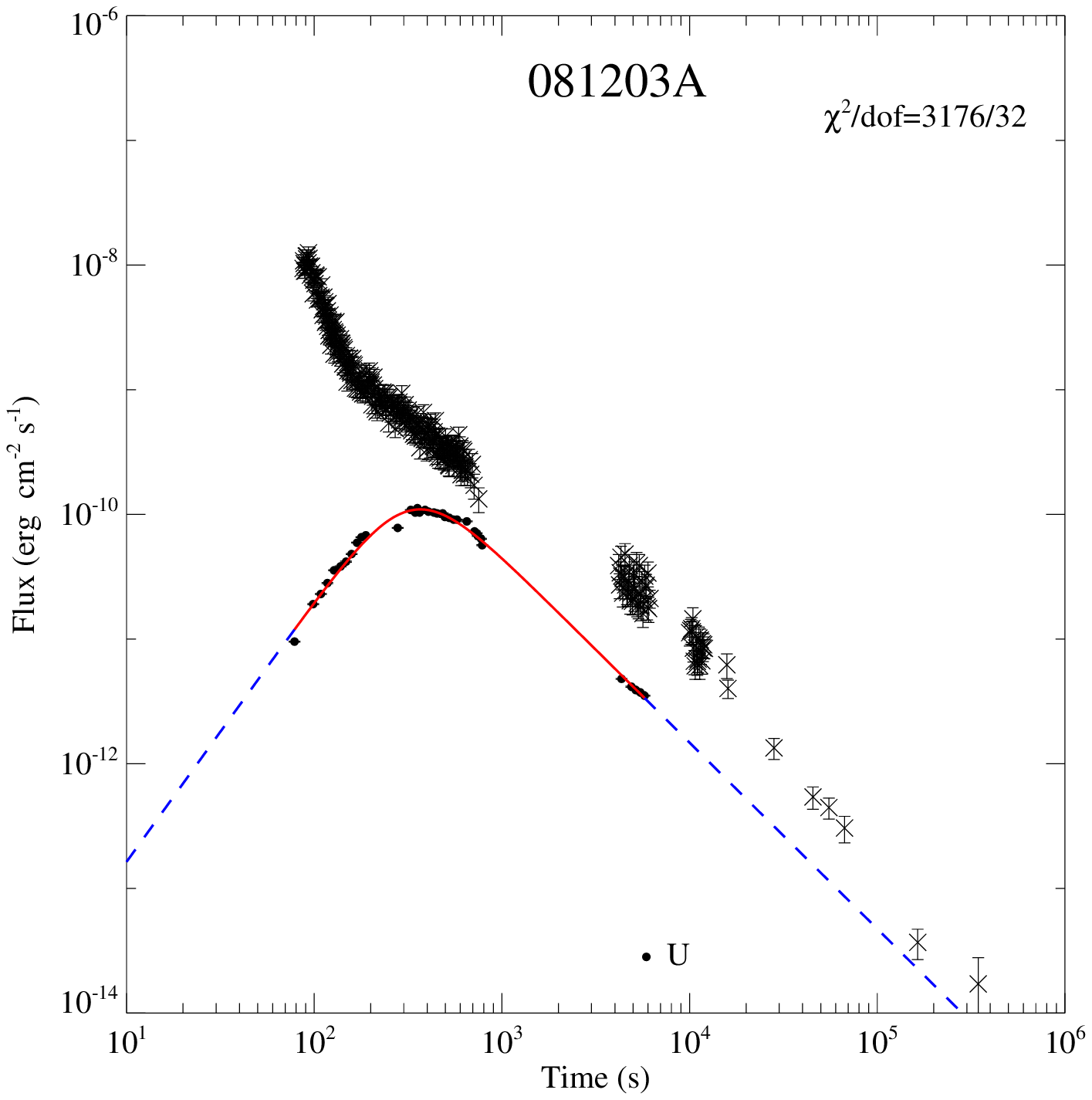}
\center{Fig. 1--- continued}
\end{figure*}

\clearpage

\begin{figure*}
\includegraphics[angle=0,scale=0.350]{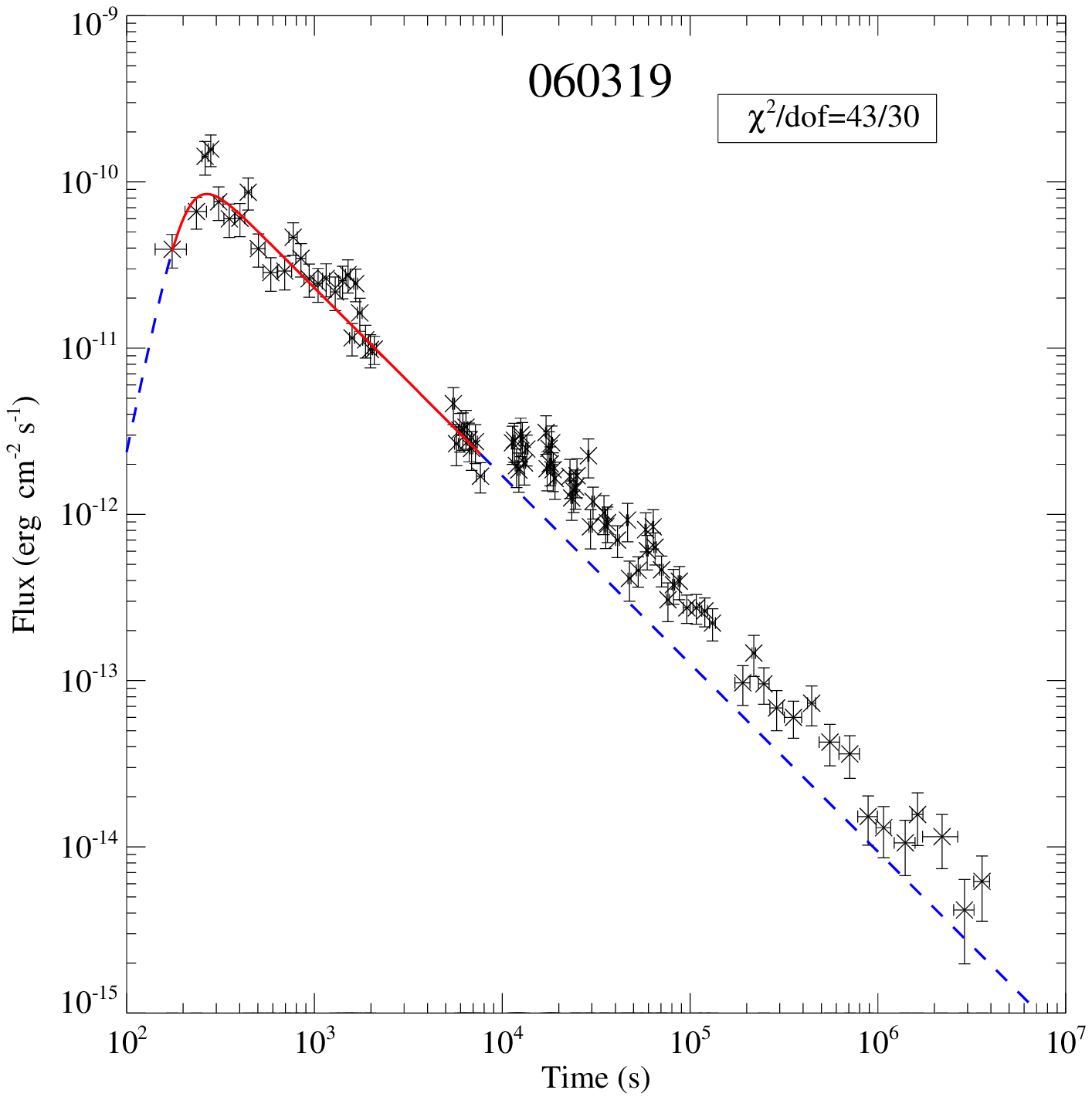}
\includegraphics[angle=0,scale=0.350]{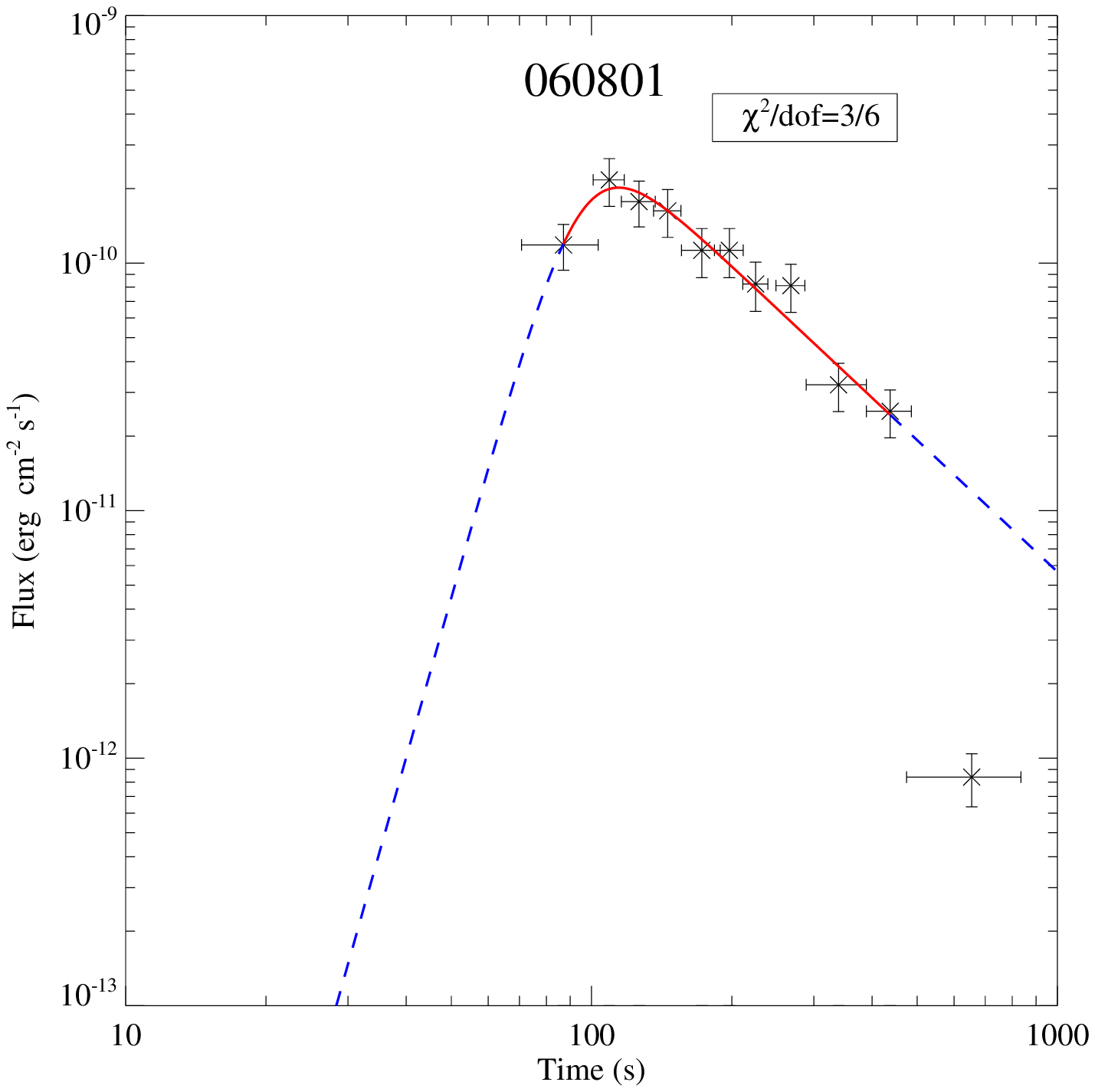}
\includegraphics[angle=0,scale=0.350]{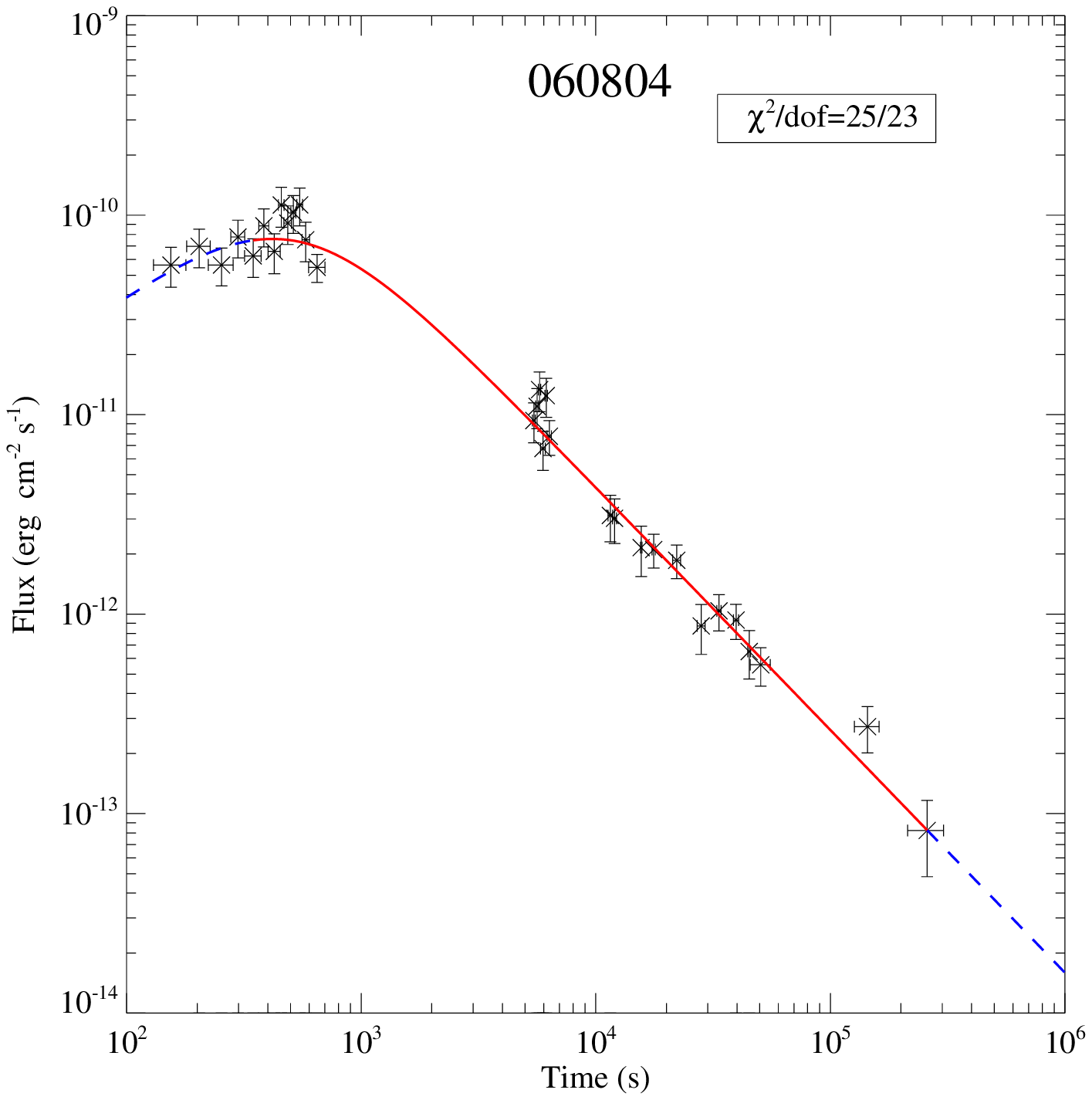}
\includegraphics[angle=0,scale=0.350]{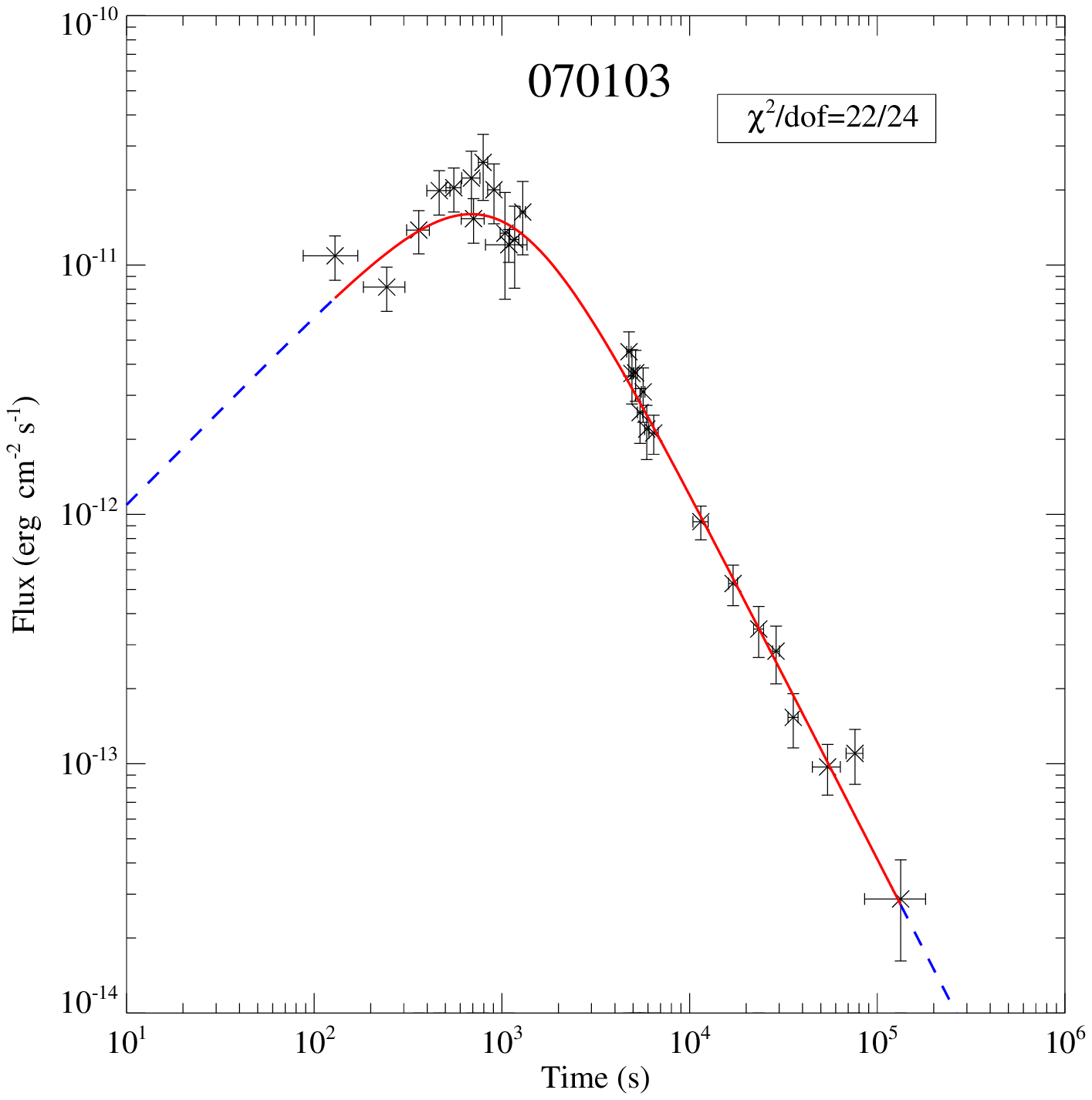}
\includegraphics[angle=0,scale=0.350]{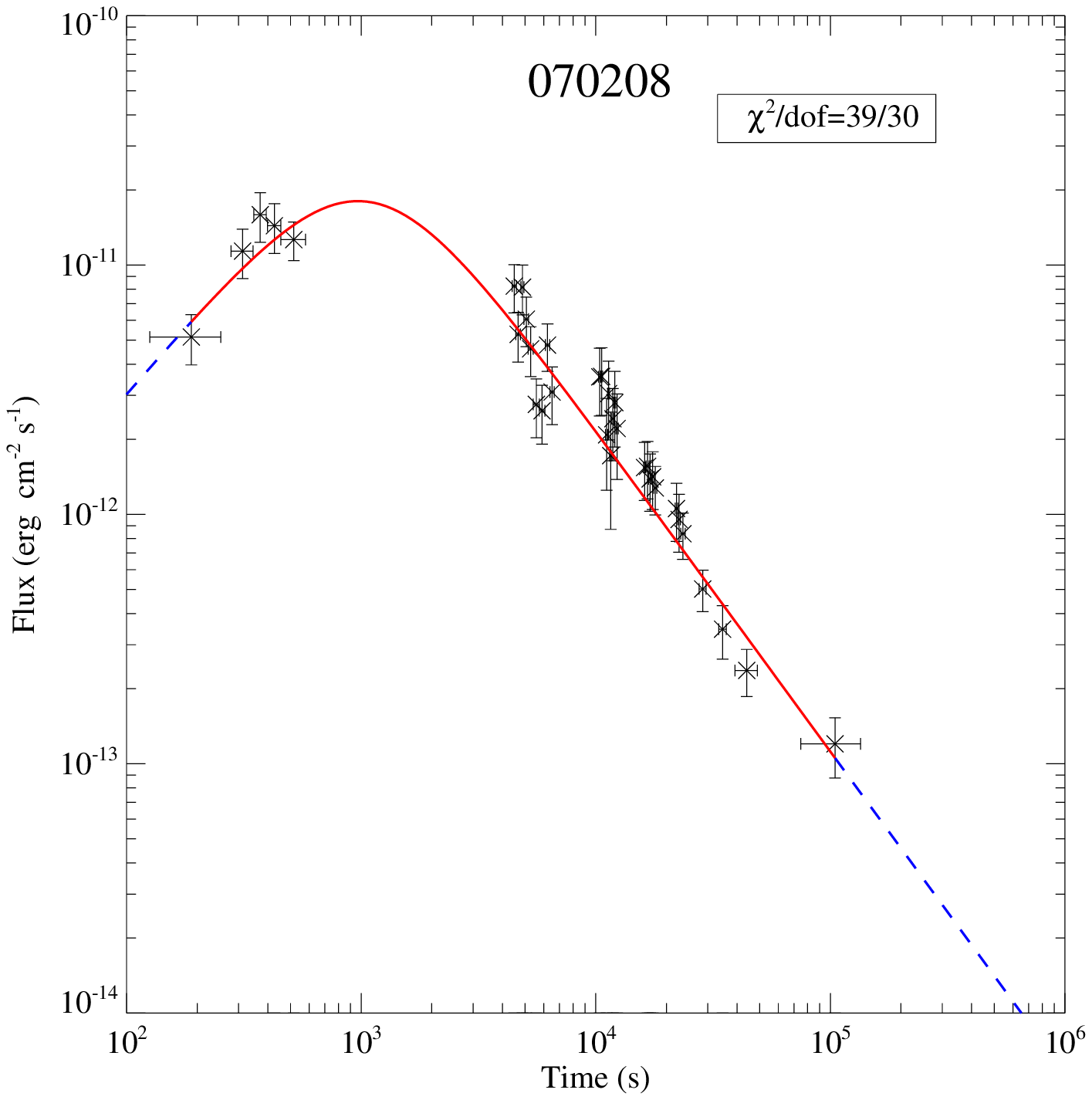}
\includegraphics[angle=0,scale=0.350]{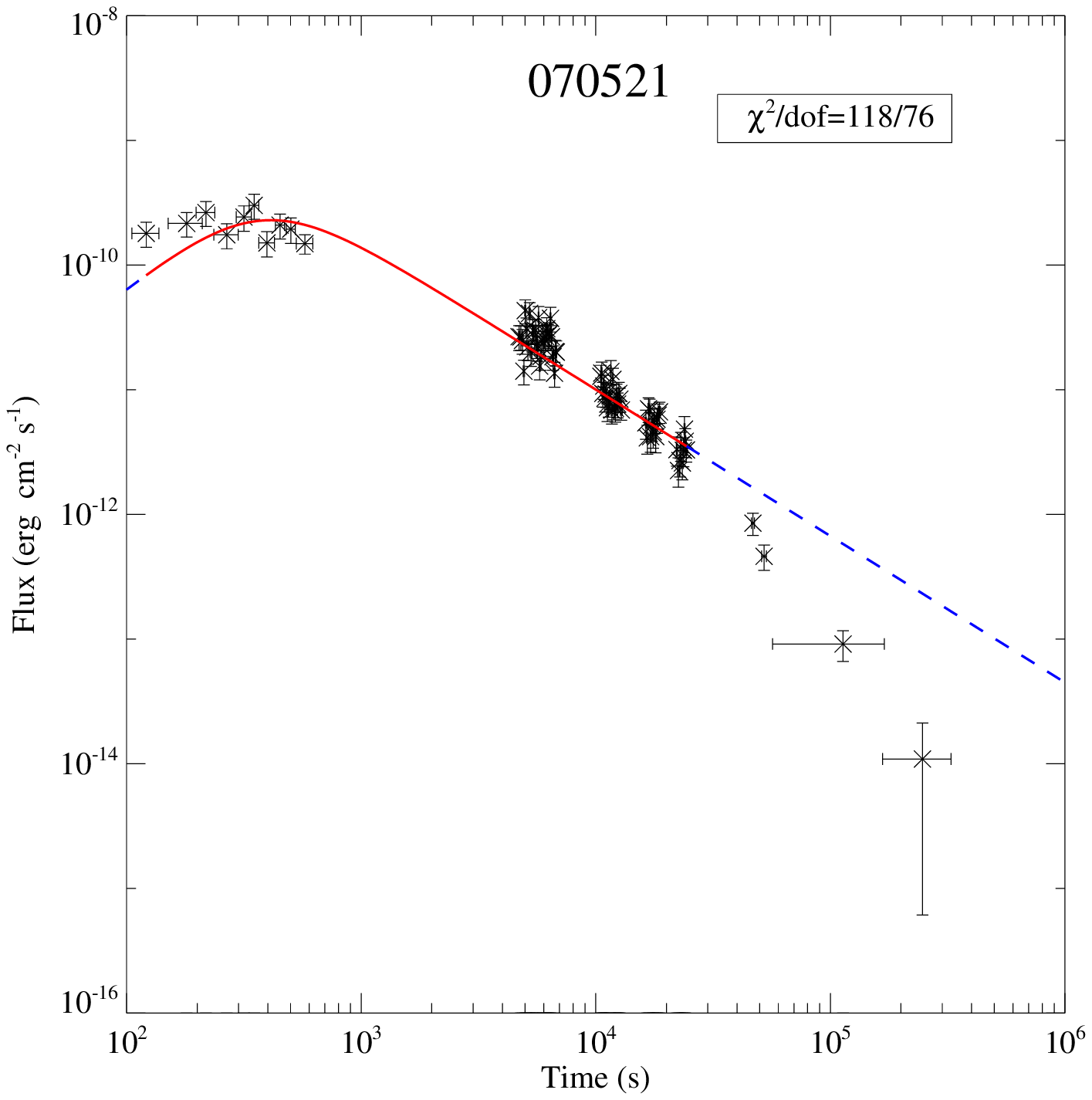}
\includegraphics[angle=0,scale=0.350]{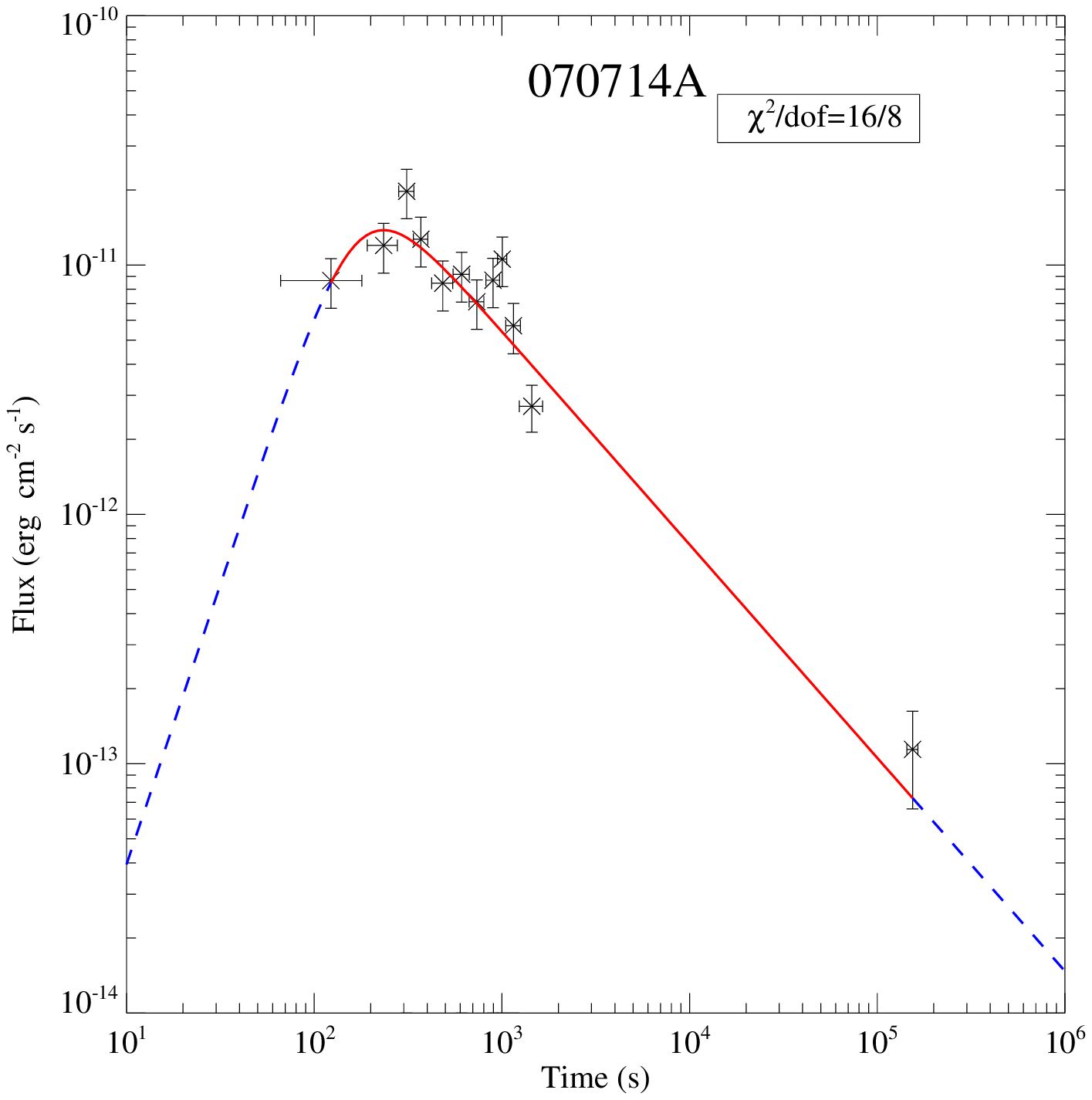}
\includegraphics[angle=0,scale=0.350]{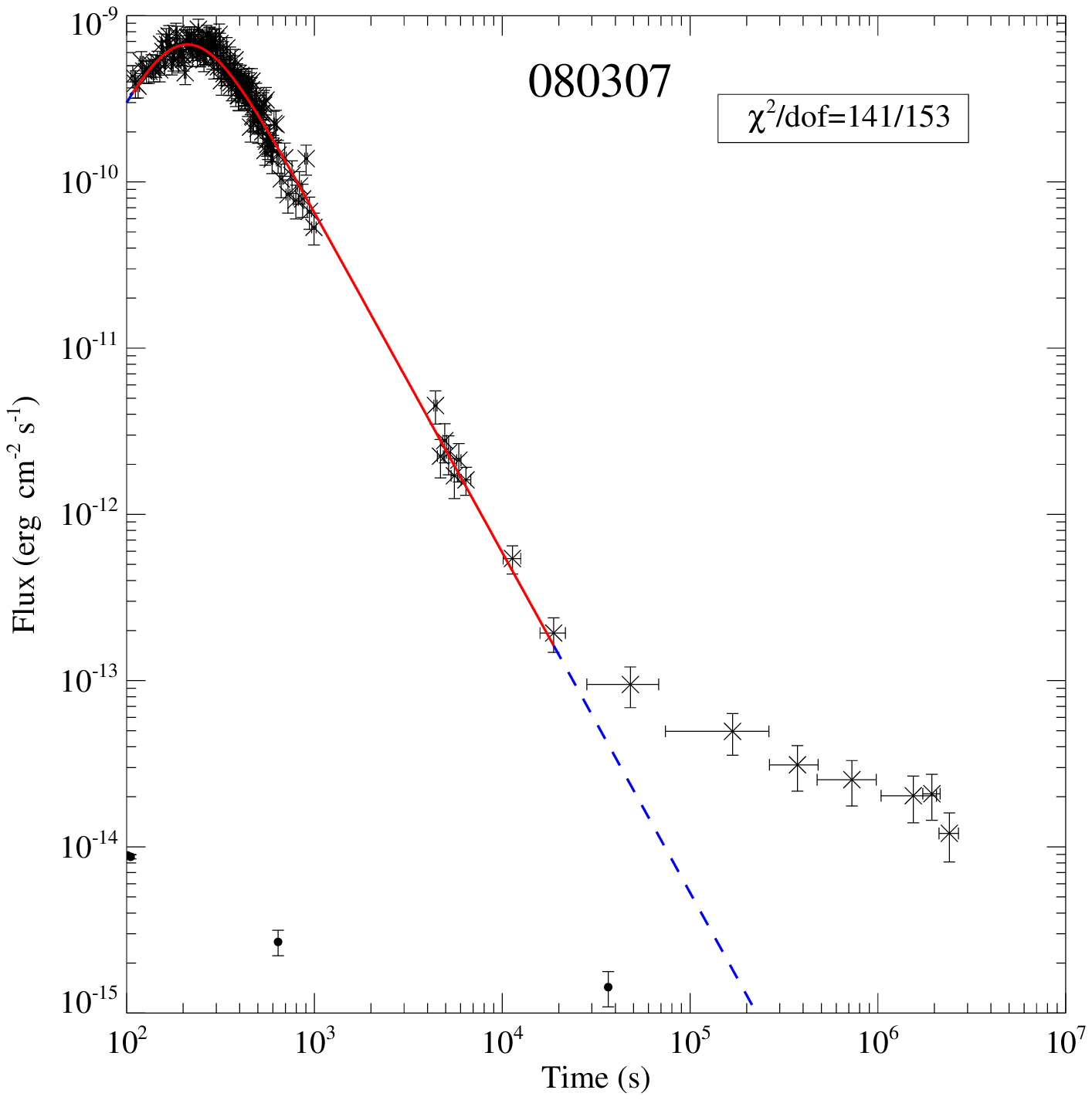}
\includegraphics[angle=0,scale=0.350]{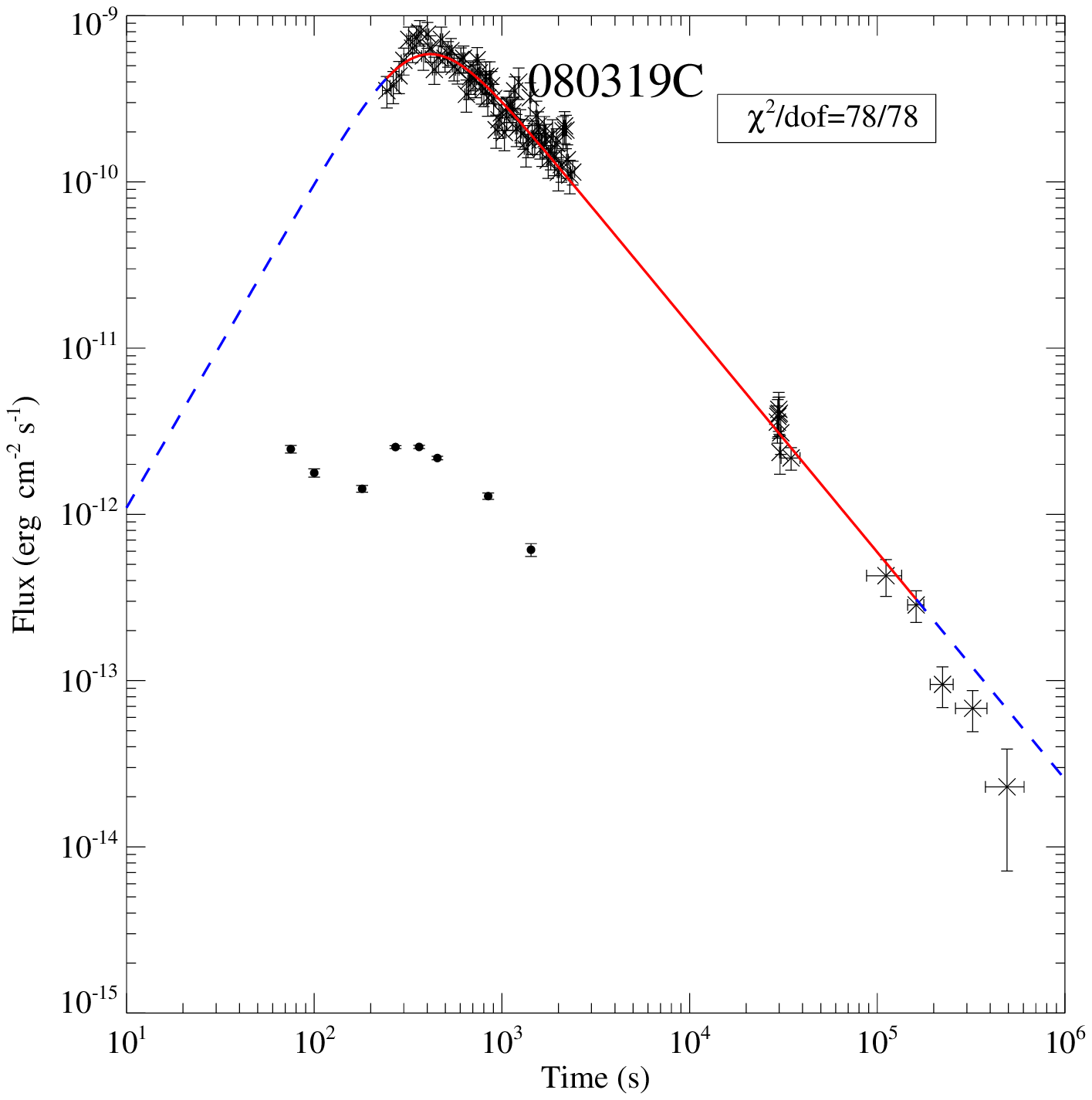}
\includegraphics[angle=0,scale=0.350]{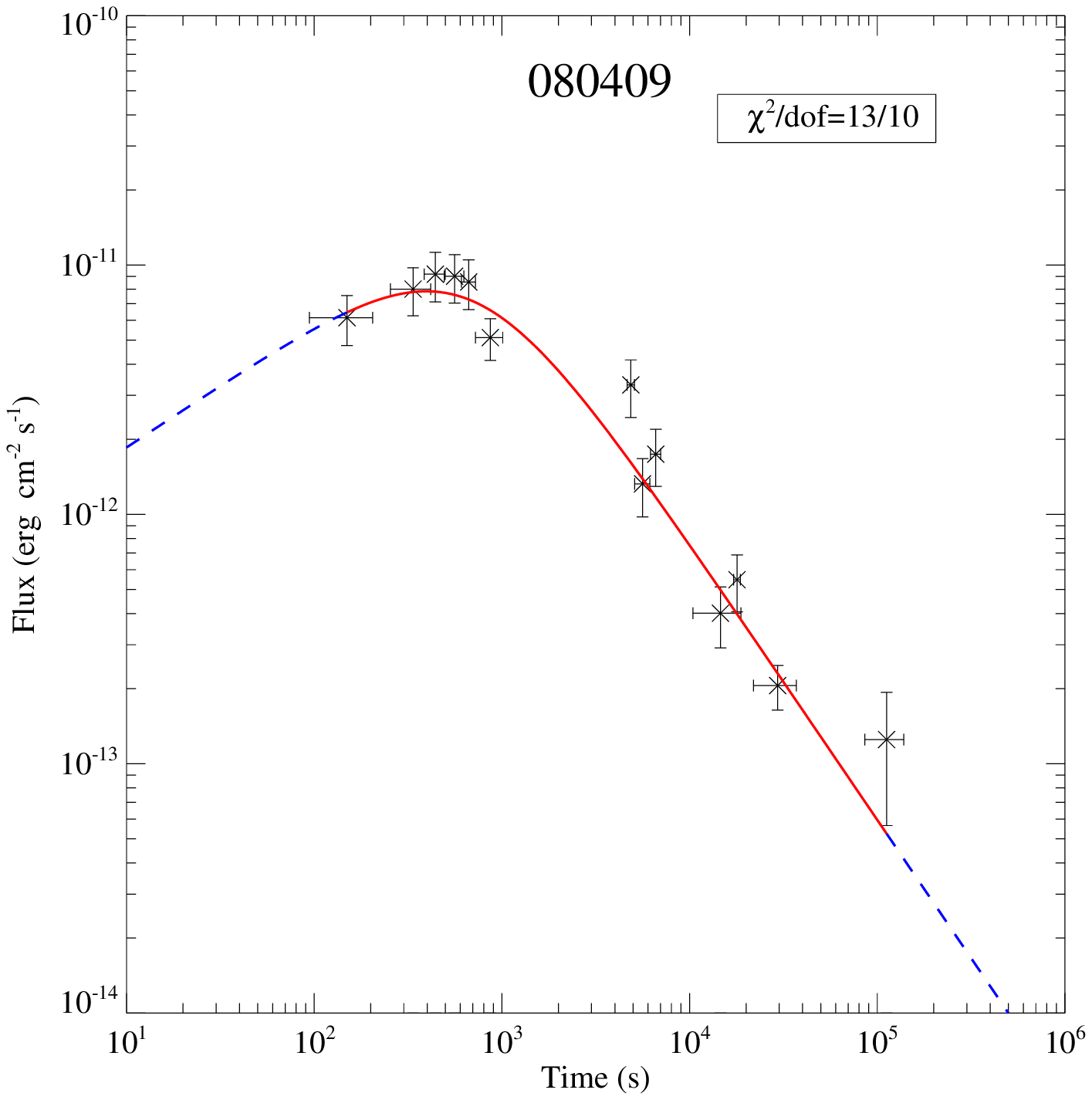}
\includegraphics[angle=0,scale=0.350]{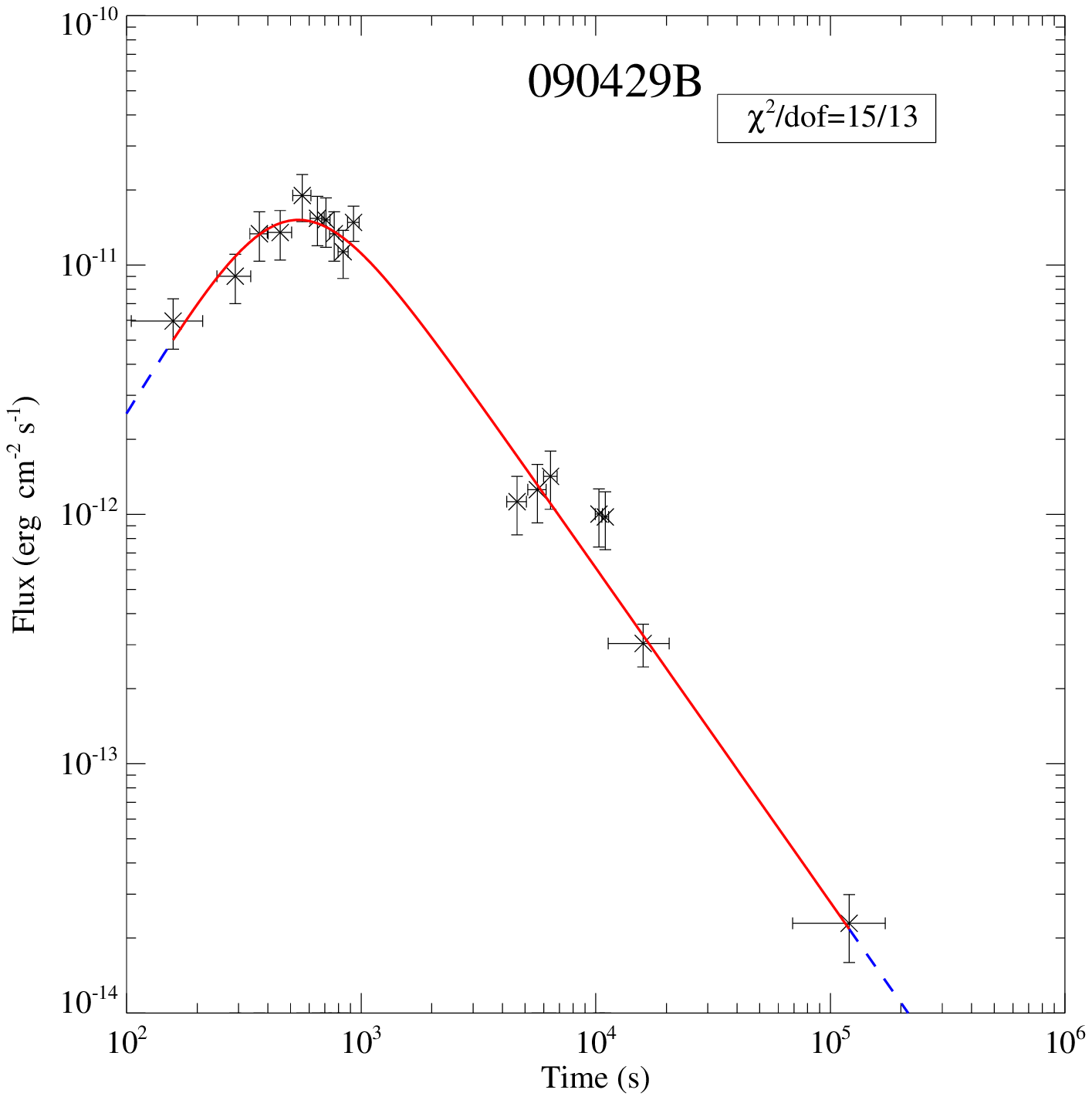}
\hfill
\includegraphics[angle=0,scale=0.350]{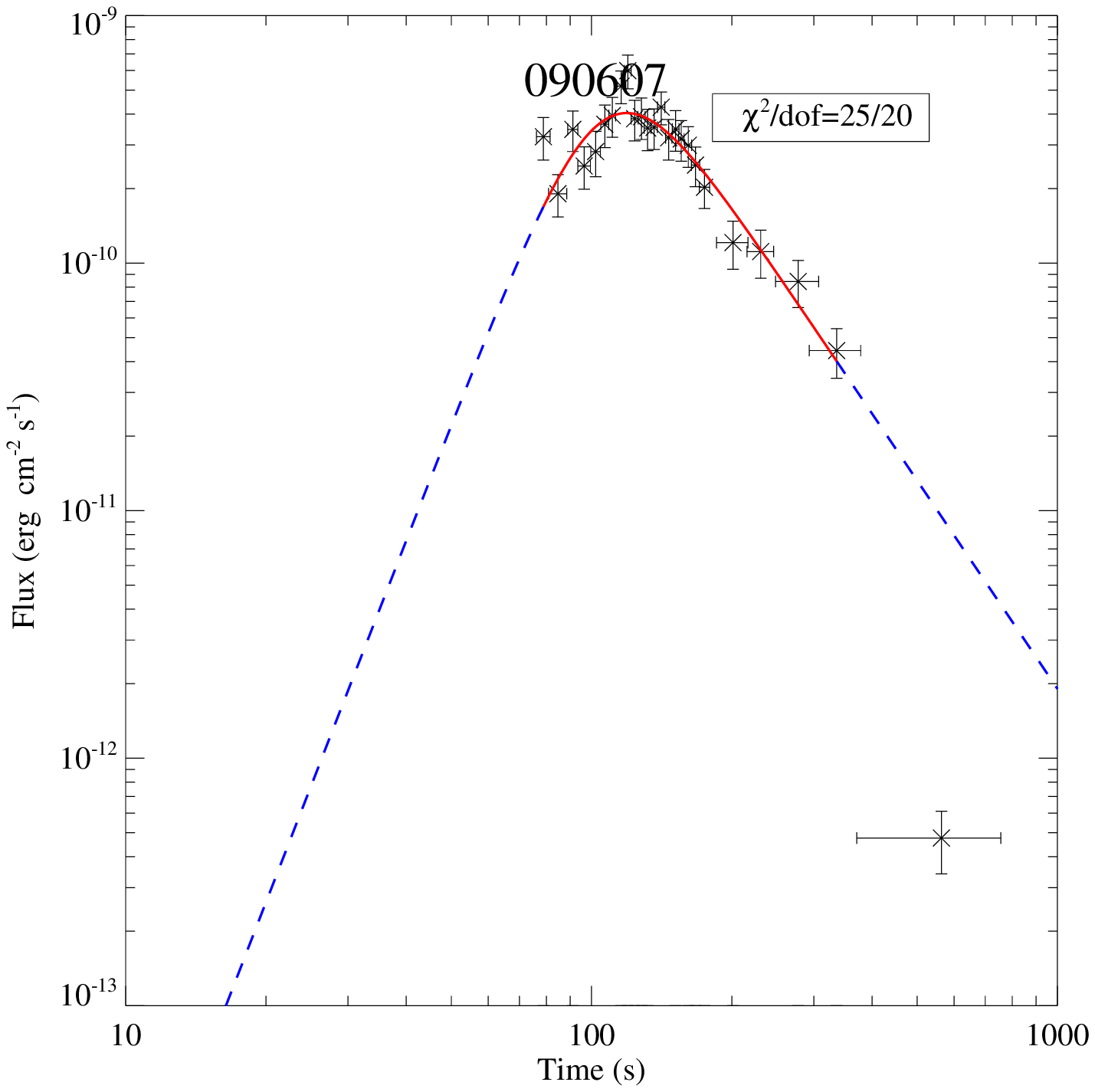}\caption{The same as Figure 1, but for the X-ray selected sample.} \label{Xray}
\end{figure*}



\clearpage \thispagestyle{empty} \setlength{\voffset}{-18mm}

\begin{figure*}
\includegraphics[angle=0,scale=0.8]{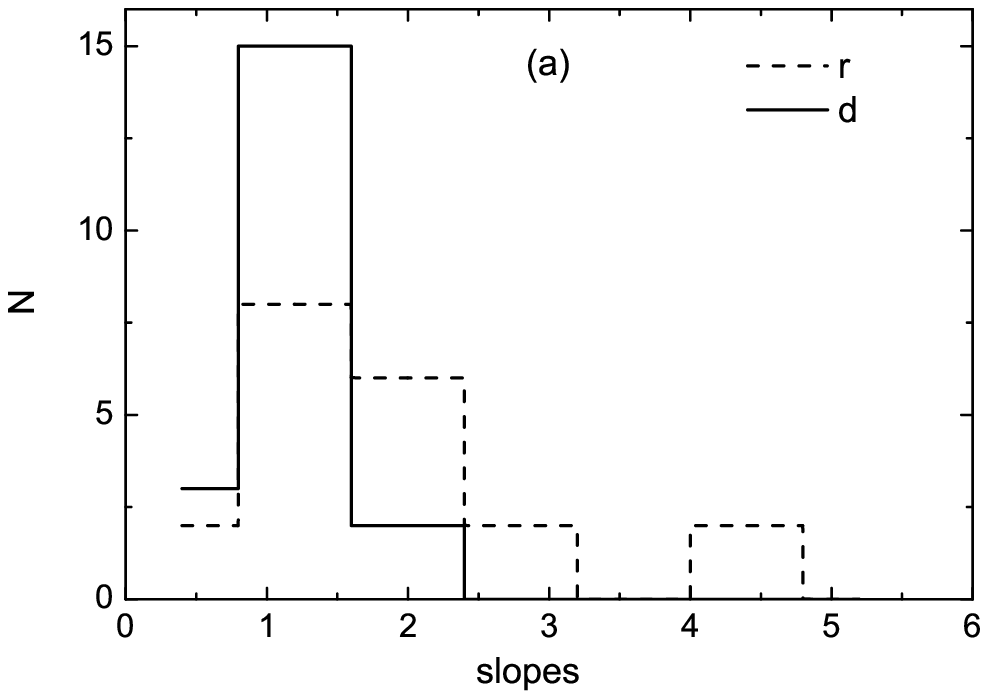}
\includegraphics[angle=0,scale=0.8]{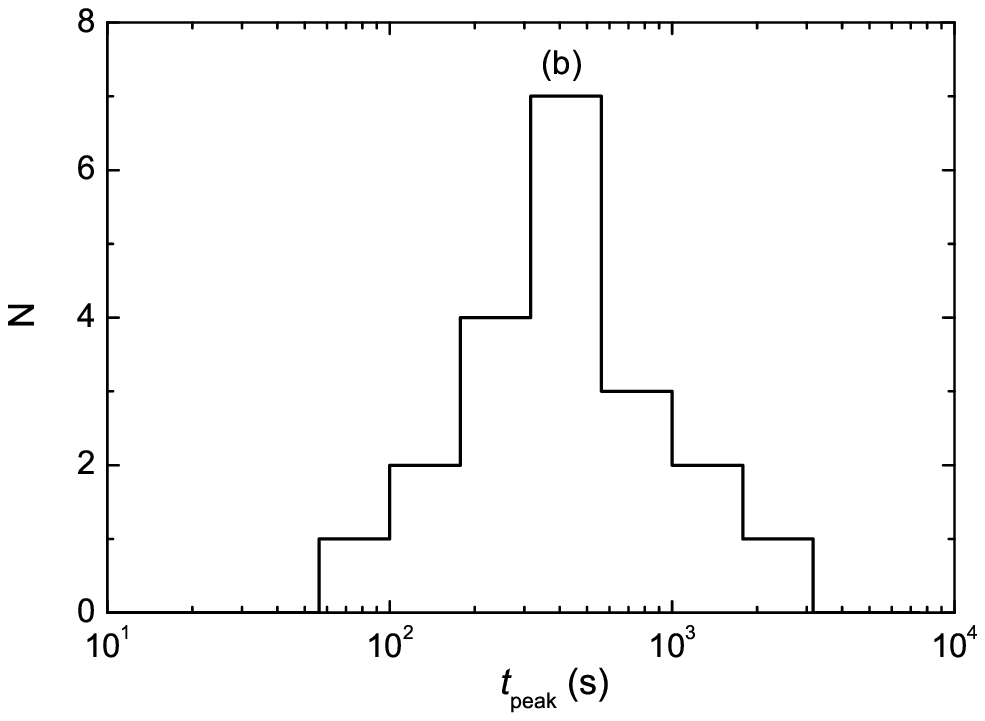}
\includegraphics[angle=0,scale=0.8]{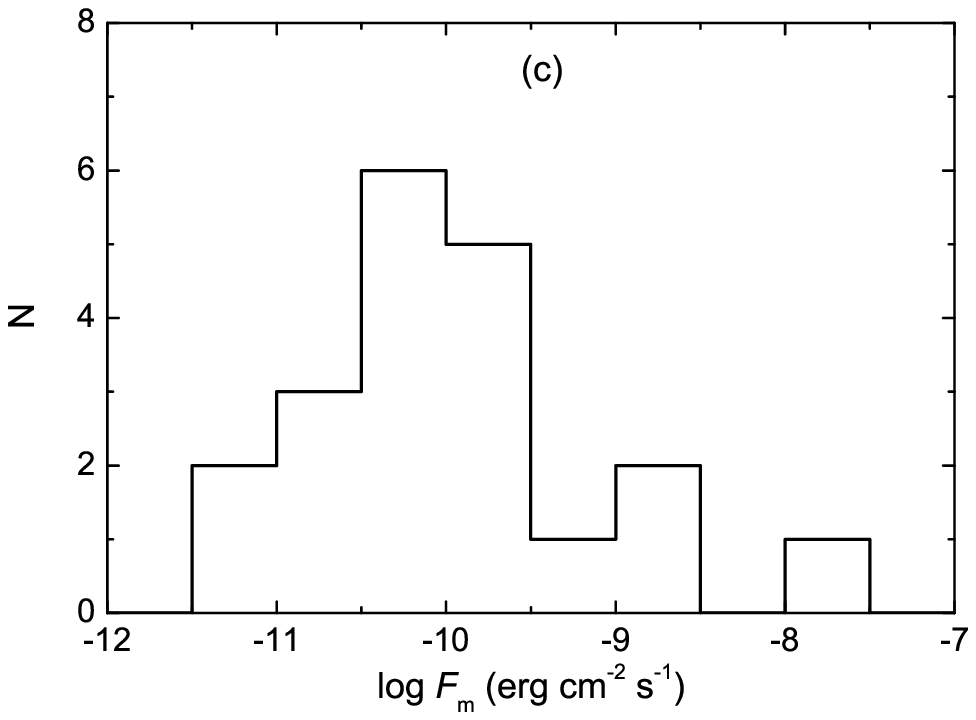}
\includegraphics[angle=0,scale=0.8]{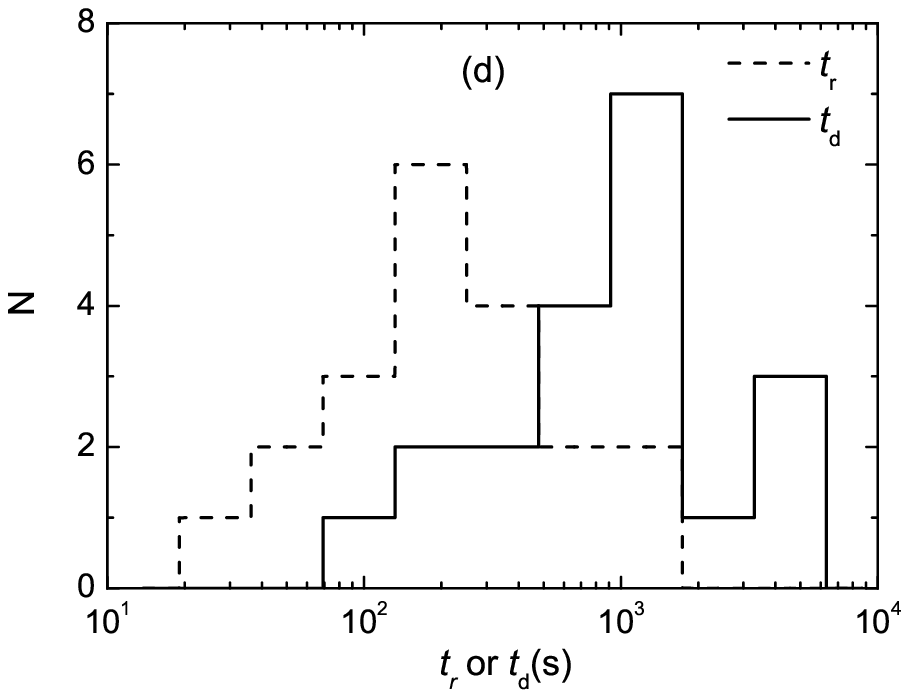}
\includegraphics[angle=0,scale=0.8]{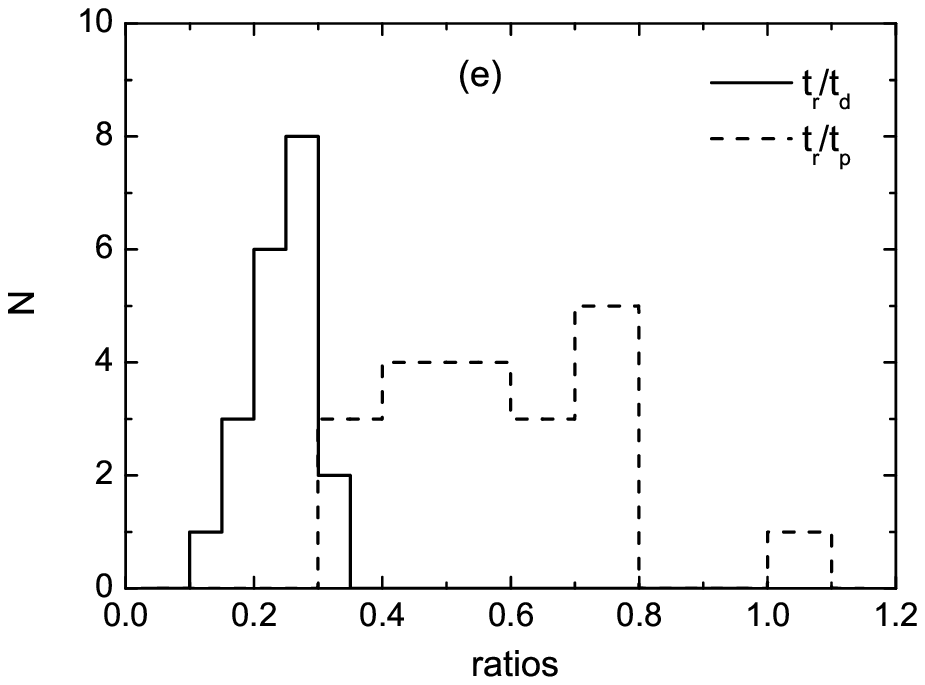}
\includegraphics[angle=0,scale=0.8]{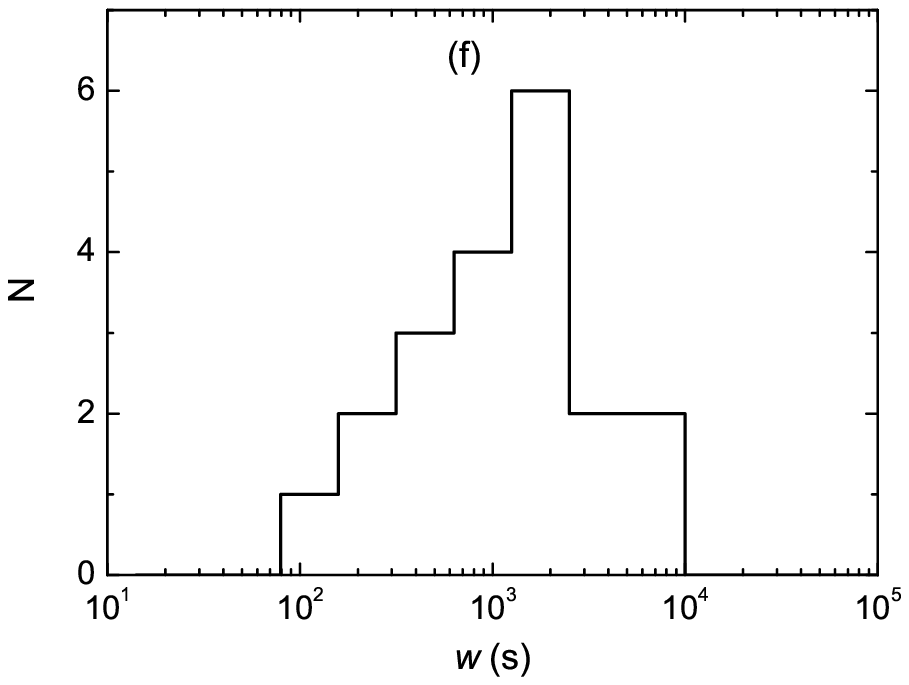}
\caption{Distributions of the characteristics of the optical-selected sample.}
\label{Dis_time}
\end{figure*}
\clearpage

\begin{figure*}
\includegraphics[angle=0,scale=0.8]{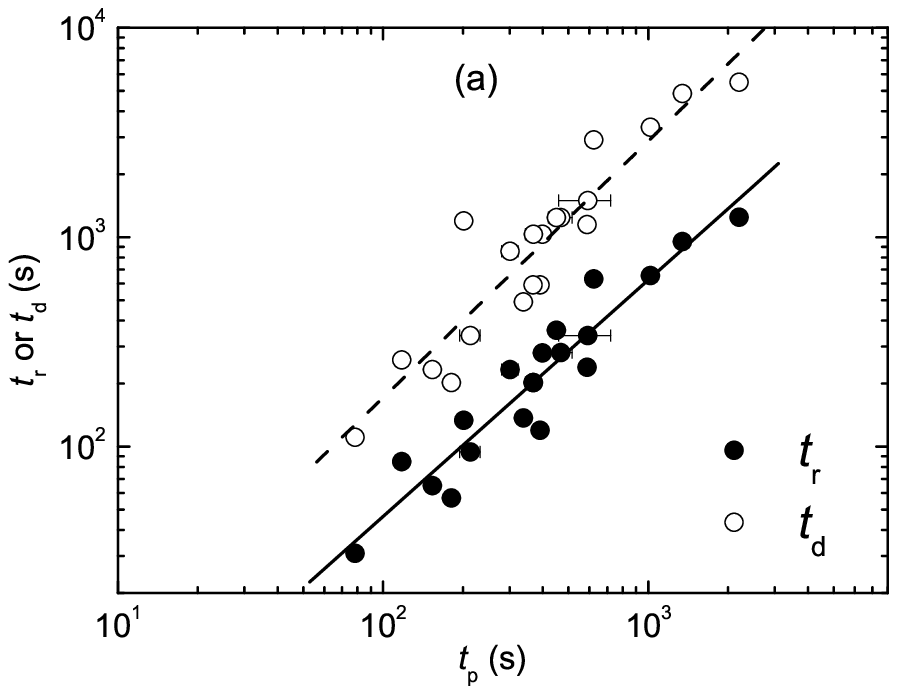}
\includegraphics[angle=0,scale=0.8]{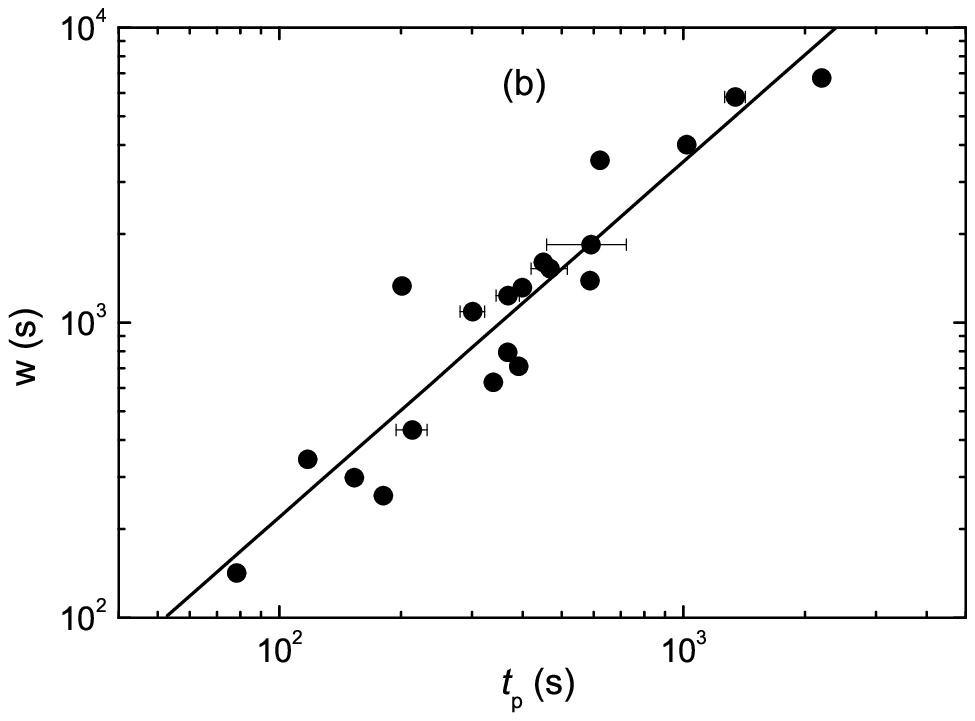}
\includegraphics[angle=0,scale=0.8]{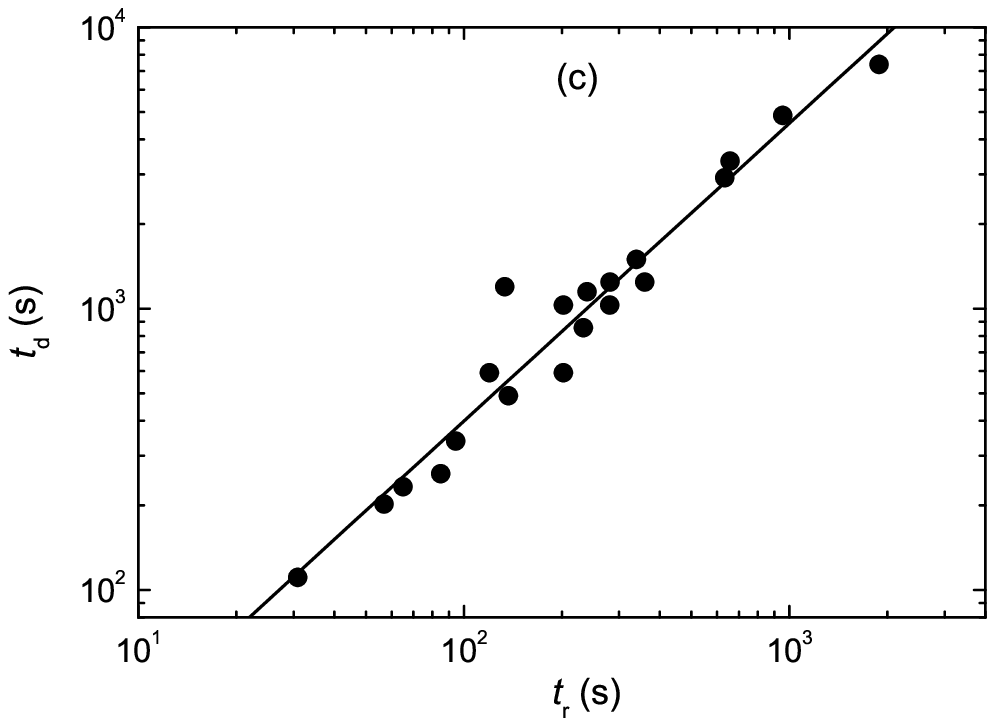}
\includegraphics[angle=0,scale=0.8]{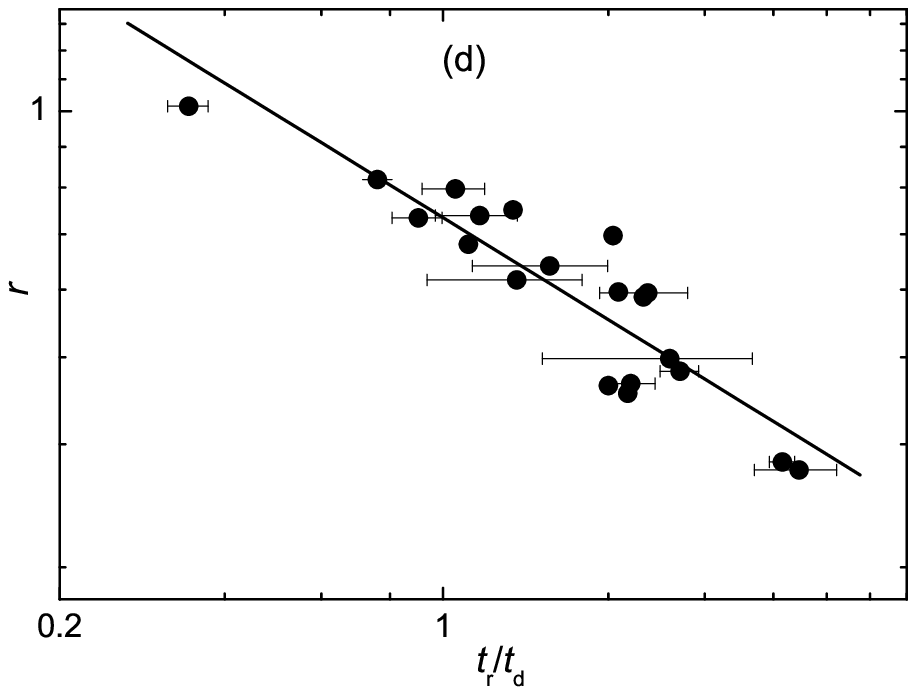}
\includegraphics[angle=0,scale=0.8]{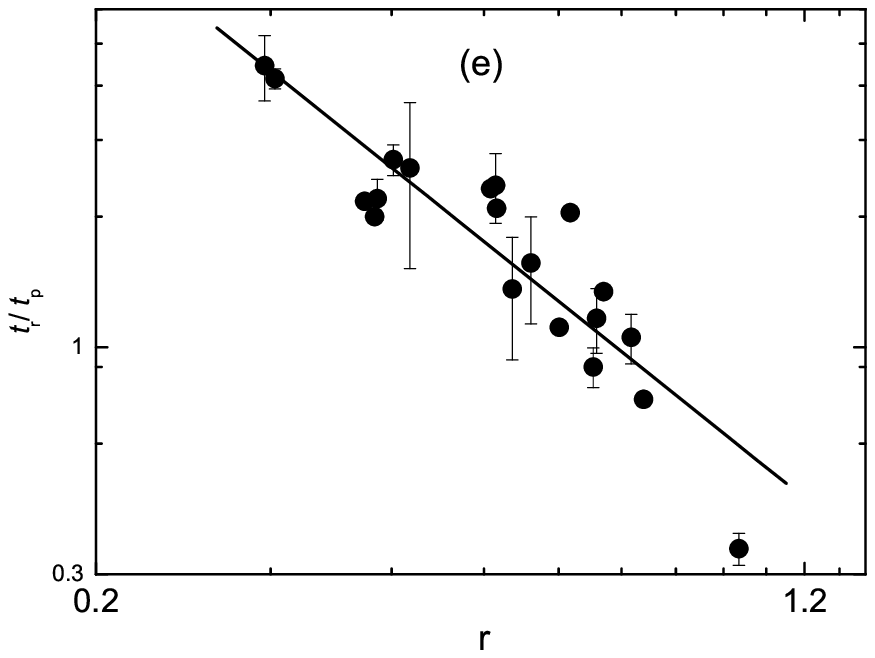}
\hfill
\includegraphics[angle=0,scale=0.8]{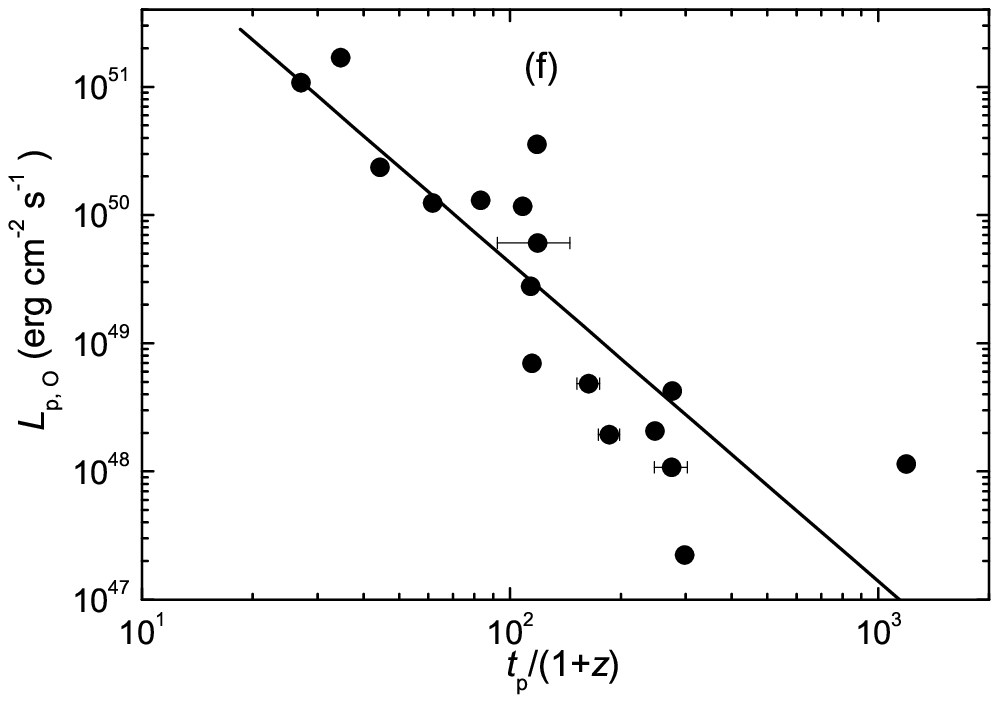}
\caption{Pair correlations among the characteristics of the optical-selected
sample. Lines are the best fits.} \label{Corr_Time}
\end{figure*}
\clearpage

\begin{figure*}
\includegraphics[angle=0,scale=0.8]{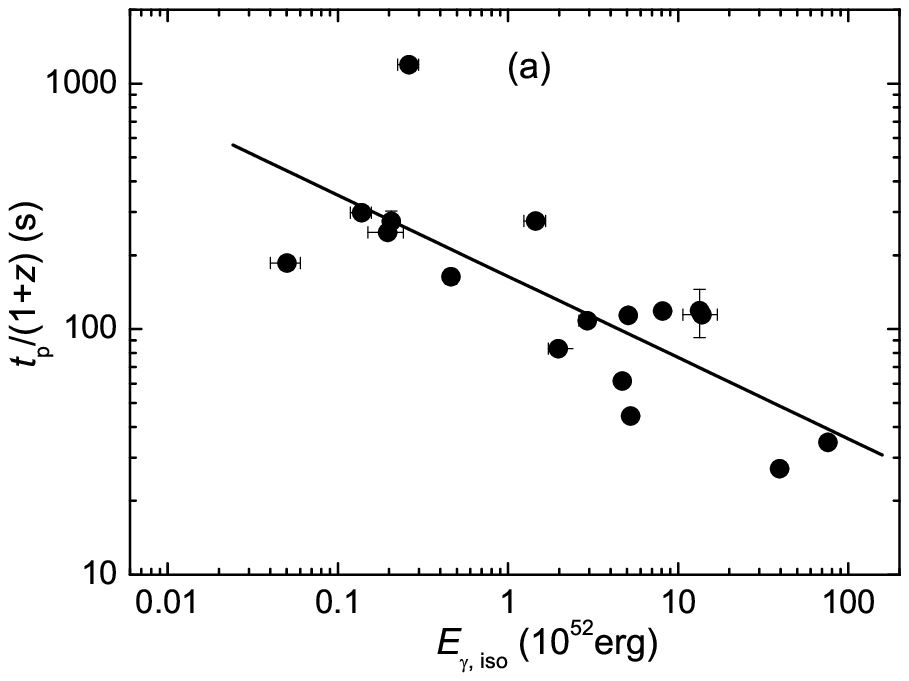}
\includegraphics[angle=0,scale=0.8]{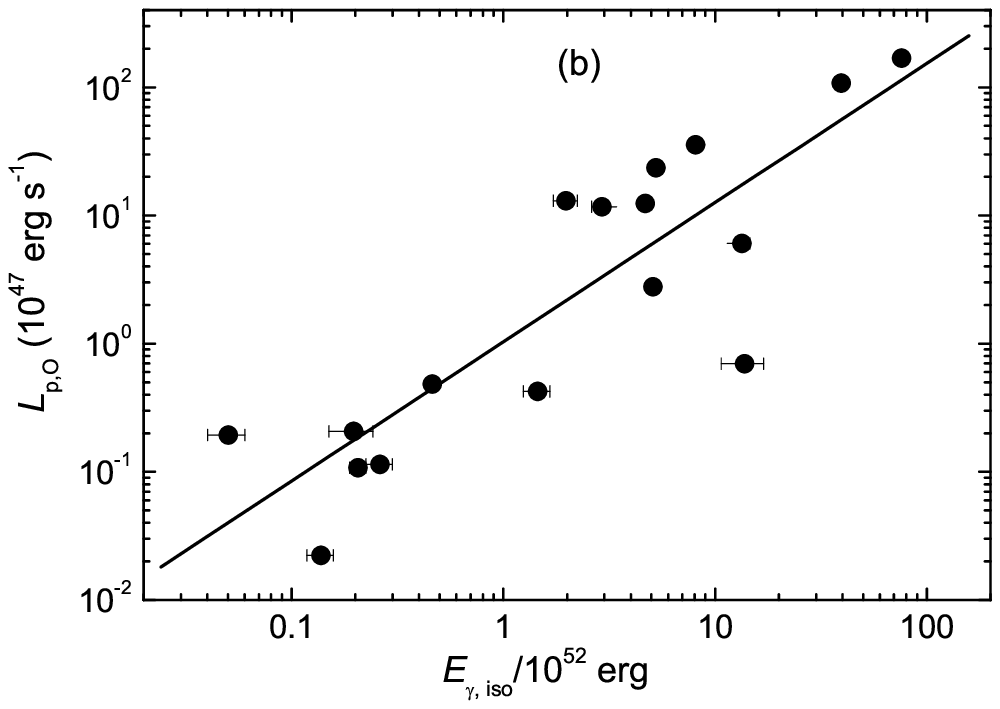}
\includegraphics[angle=0,scale=0.8]{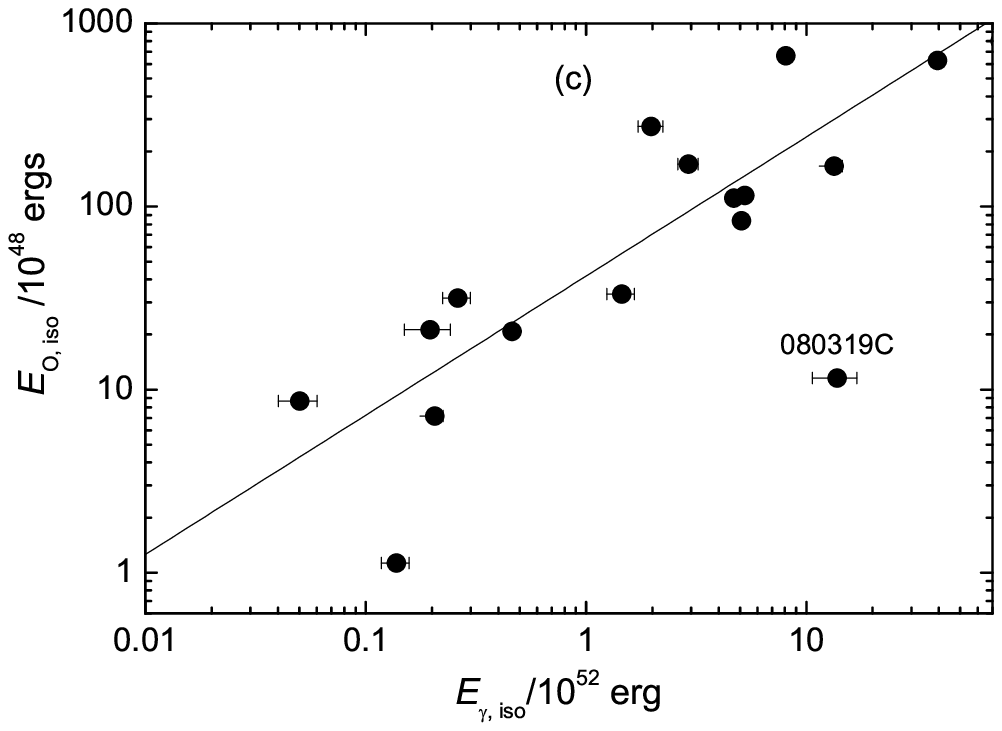}
\caption{Relations of $E_{\rm iso, \gamma}$ to $t_p/(1+z)$, $L_{p,O}$ and
$E_{\rm iso, O}$ for the optical selected sample. Lines are the best fits.}
\label{Corr_Eiso_Lp}
\end{figure*}

\begin{figure*}
\includegraphics[angle=0,scale=0.8]{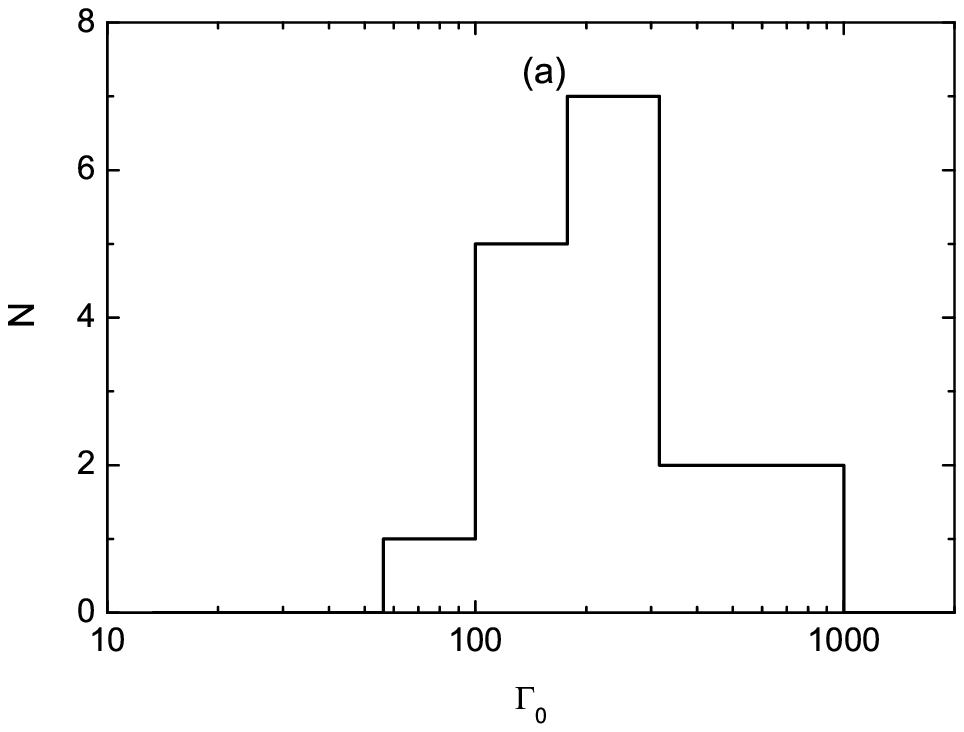}
\includegraphics[angle=0,scale=0.8]{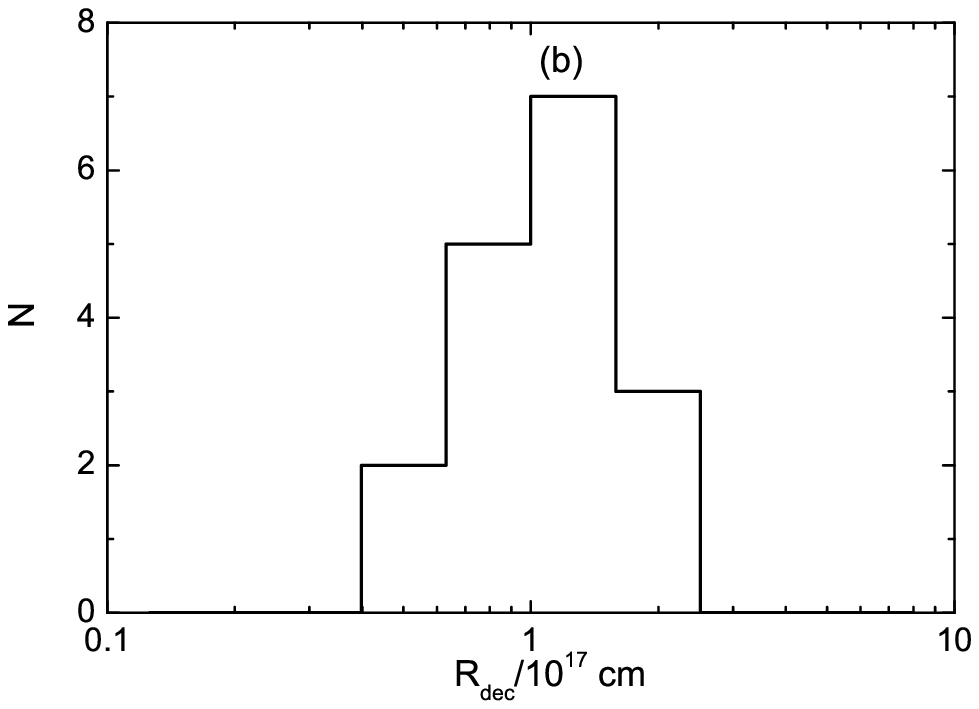}
\caption{Distributions of $\Gamma_0$ and $R_{d}$ for the optical-selected
sample} \label{Dis_Gamma}
\end{figure*}

\begin{figure*}
\includegraphics[angle=0,scale=0.8]{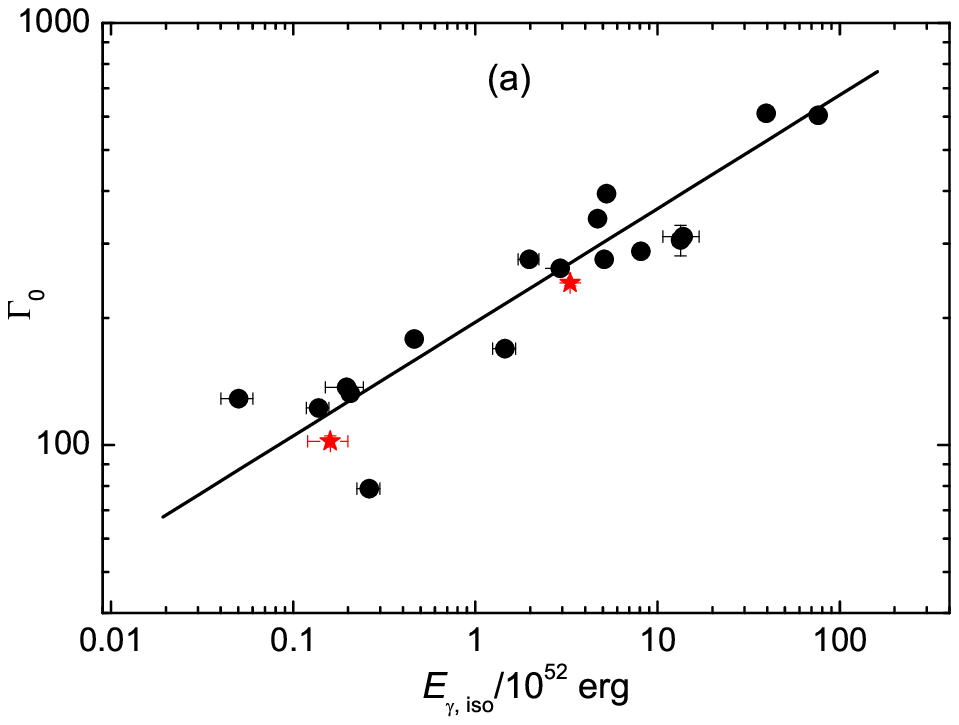}
\includegraphics[angle=0,scale=0.8]{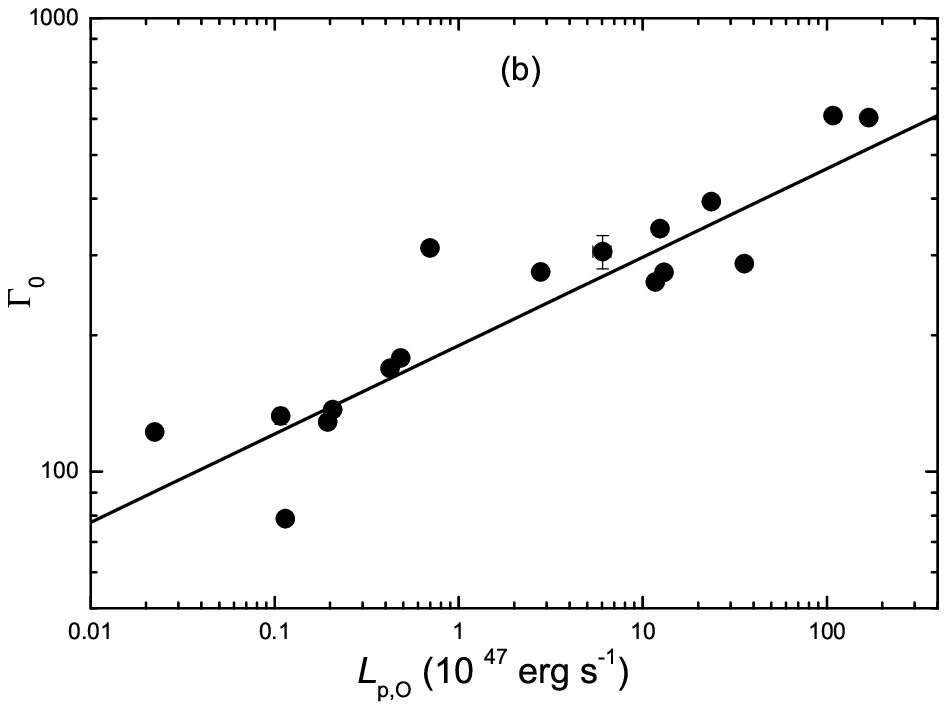}
\hfill
\includegraphics[angle=0,scale=0.8]{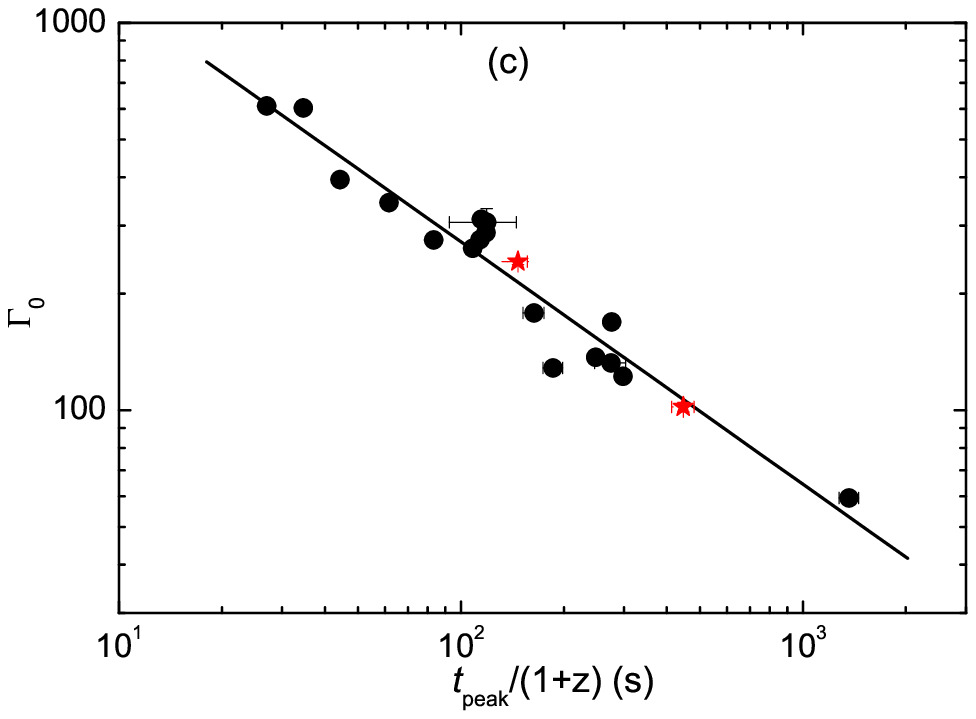}
\caption{Relations of $\Gamma_0$ to $E_{\rm iso,\gamma}$, $L_{\rm p,O}$, and
$t_{p}/(1+z)$. Lines are the best fits for the optical-selected sample. GRBs
070208 and 080319C in the X-ray selected sample have redshift measurement. We
mark the two GRBs with red stars.} \label{Corr_Gamma}
\end{figure*}
\end{document}